\documentclass[sigconf, review = false]{acmart}
\AtBeginDocument{%
  }

\acmDOI{}
\acmISBN{}
\acmYear{}
\setcopyright{none}
\acmConference[]{}{}{}
\usepackage[yyyymmdd]{datetime}

\usepackage{csquotes}
\usepackage{threeparttable}

\usepackage{pgfplots}
\pgfplotsset{compat=1.18}
\usetikzlibrary{pgfplots.statistics}

\usepackage{tabularx}
\usepackage{dcolumn} 
\newcolumntype{d}[1]{D{.}{.}{#1}}

\usepackage[dvipsnames]{xcolor}
\usepackage[most]{tcolorbox}

\usepackage{subcaption}
\usepackage{enumitem}
\usepackage{listings}
\usepackage{makecell}

\usepackage{todonotes}
\let\xtodo\todo
\renewcommand{\todo}[1]{\xtodo[inline,color=green!50]{#1}}

\begin{document}


\title{From Semantic Editing Intentions to Persistent Parametric Controls for AI-Assisted Image Editing}

\author{Rifat Mehreen Amin}
\orcid{0000-0003-4279-7778}
\affiliation{%
  \institution{LMU Munich}
  \city{Munich}
  \country{Germany}
}
\email{rifat.amin@ifi.lmu.de}

\author{Michelle Dutoit}
\orcid{0009-0006-7926-6873}
\affiliation{%
  \institution{LMU Munich}
  \city{Munich}
  \country{Germany}
}
\email{michelle.dutoit@ifi.lmu.de}

\author{Anton Rockenstein}
\orcid{0009-0009-5029-1575}
\affiliation{%
  \institution{LMU Munich}
  \city{Munich}
  \country{Germany}
}
\email{anton.rockenstein@campus.lmu.de}

\author{Baptiste Caramiaux}
\orcid{0000-0002-9007-9888}
\affiliation{%
 \institution{CNRS, Sorbonne Université}
  \city{Paris}
  \country{France}}
\email{caramiaux@isir.upmc.fr}

\author{Andreas Butz}
\orcid{0000-0002-9007-9888}
\affiliation{%
 \institution{LMU Munich}
  \city{Munich}
  \country{Germany}}
\email{butz@ifi.lmu.de}

\renewcommand{\shortauthors}{Amin et al.}

\begin{abstract}

AI-powered image editing increasingly allows users to express high-level editing intentions, yet how these intentions are translated into concrete editing actions often remains difficult to understand. We present Atelier, a semantic image editing tool that uses image semantics and user intent to determine which editing controls to surface and how to initialize them. Guided by a formative study with eight image-editing practitioners, Atelier composes context-specific, directly manipulable controls that users can refine rather than committing AI interpretations directly to finished edits. In a within-subjects study with 12 practitioners, participants preferred AI-composed controls, described them as supporting creative exploration, and actively refined AI proposals. However, experiences of control were mixed, and interviews highlighted limited insight into why particular controls and values were proposed. Our findings suggest that making AI interpretations actionable through manipulable controls can support exploration and refinement, but actionability alone does not make those interpretations transparent to all users.

\end{abstract}



\begin{teaserfigure}
 \includegraphics[width=\textwidth]{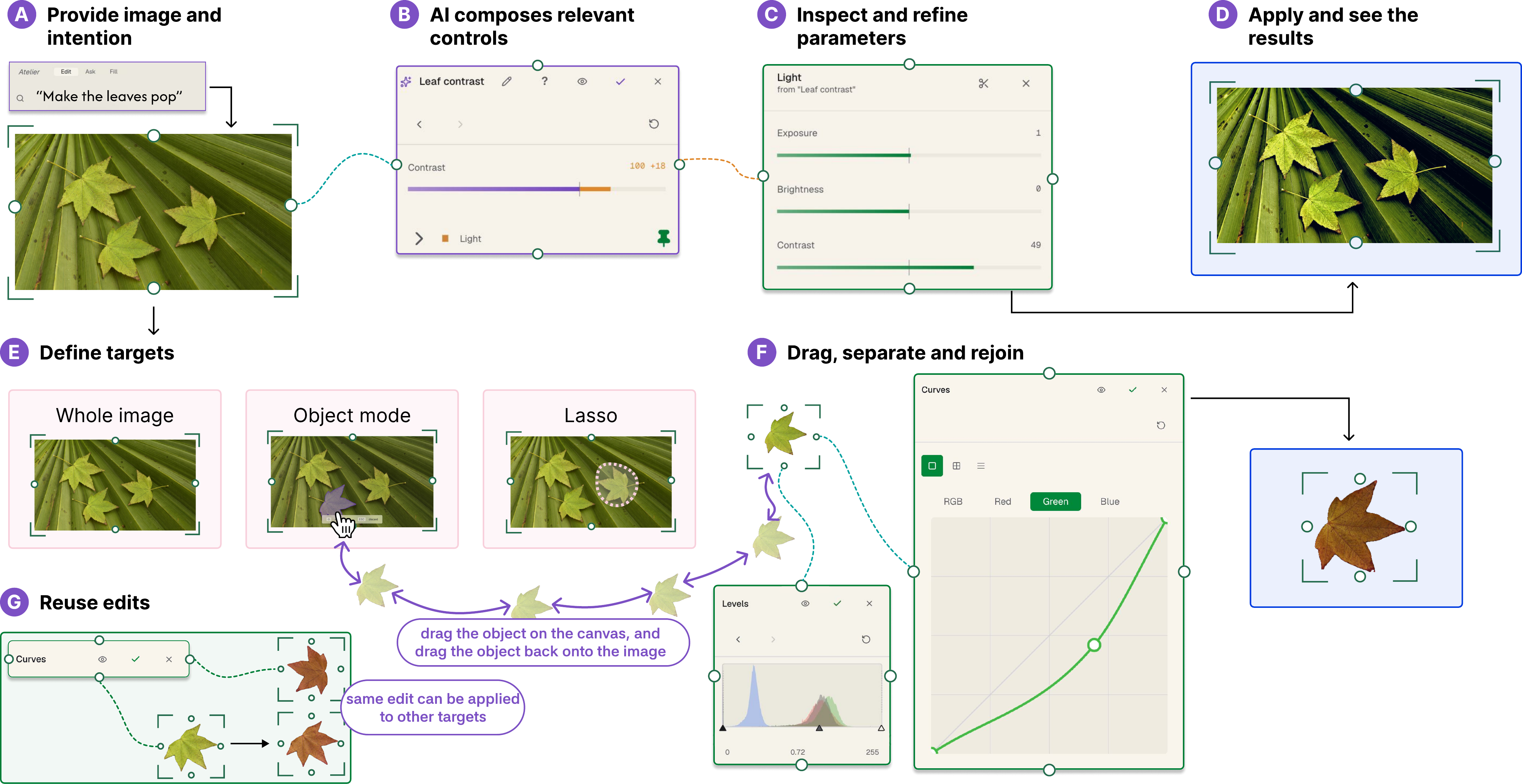}
\caption{
\textbf{Atelier: translating image-editing intentions into persistent, manipulable parametric controls.}
Users can (A) provide an image and a high-level editing intention,
(B) receive context-specific controls composed from the image and intention,
(C) inspect and refine the underlying editing parameters, and (D) immediately evaluate their effect on the image. Users can also (E) scope edits to the whole image, selected objects, or lasso-defined regions,
(F) drag and separate persistent targets and controls on the canvas, and rejoin the separated targets back to the source image after editing, and (G) reuse an existing edit by applying the same control configuration to additional targets.
}
\Description{Seven-part diagram of Atelier's editing workflow: a top pipeline (A–D)
and three further interaction patterns (E–G). 
A shows an image of three maple leaves with the typed intention ``Make the leaves pop.'' B shows a connected Leaf contrast widget: a high-level contrast slider above an underlying Light operation. C shows the Light operation expanded into three sliders: exposure, brightness, and contrast. D shows the resulting image, with stronger contrast and more visually prominent leaves, marked with a blue border as the final result. 
E shows three panels illustrating ways to scope an edit: Whole image, with no selection marks; Object mode, showing a cursor clicking on one leaf; and Lasso, showing a dashed circular outline drawn around part of a leaf.
F shows a leaf object dragged out of the original image and connected by dashed lines to a Levels widget and a Curves widget; an arrow leads to the edited leaf now shifted to a browned, aged tone and marked with a blue border as the result. G shows a Curves control connected to one leaf image, with an arrow to two further leaf thumbnails that have taken on the same browned, aged tone, and an on-screen callout stating that the edit can be applied to other targets.
}
  \label{fig:teaser}
\end{teaserfigure}

\received{20 February 2007}
\received[revised]{12 March 2009}
\received[accepted]{5 June 2009}

\maketitle

\section{Introduction}

Recent advances in multimodal foundation models and generative image models have expanded image editing toward semantic, instruction-driven interaction, allowing users to express desired changes to objects, regions, styles, and other high-level visual attributes \cite{partedit,promptpaint,wang_promptcharm_2024}. However, expressing \emph{what} should change still leaves the system to determine \emph{how} that intention should be translated into concrete edits. Making an image ``more dramatic,'' for example, might involve changes to contrast, lighting, color, or specific regions. \citet{subramonyam_gulf_2024} describe a related challenge as the \emph{gulf of envisioning}: the cognitive distance between a user's intention and an instruction that anticipates how an AI system will realize it. This challenge is particularly relevant in creative work, where intentions often develop through exploration. When an AI's interpretation is expressed only through an edited result, users can evaluate the outcome but have limited opportunity to engage with the decisions that produced it. We therefore investigate how these interpretations can instead become manipulable parts of the editing interface.

Prior work has helped users formulate editing instructions \cite{promptify,wang_promptcharm_2024}, specify changes through visual and multimodal interaction \cite{promptpaint,visualprompts}, and manipulate generative outcomes through graphical controls \cite{davis_fashioning_2024,parkinteraction}. Generative and malleable interfaces further demonstrate how AI can construct interaction elements at runtime \cite{vaithilingam_dynavis_2024,masson_directgpt_2024, cao2025malleable,park2026semanticguidance}. We investigate a complementary question: \textbf{how can AI use the content of an image and a user's intention to decide what should become directly manipulable?} A formative study with eight image-editing practitioners showed that participants often began with semantic goals before translating them into technical parameters, worked with localized image regions, and valued AI assistance while wanting to retain the ability to refine and override its suggestions.

Guided by these findings, we present \textbf{Atelier}, a canvas-based image editor that uses visual context and user intent to compose context-specific editing controls. Drawing on the notion of \emph{interaction substrates} \cite{mackay2025substrates}, Atelier treats images and editable regions as persistent places where controls can be situated and refined. A vision-language model interprets the current image and the user's intention to identify relevant editing dimensions and spatial targets, and translates these into familiar parametric controls with proposed initial values. Because these controls use the same operations as Atelier's conventional manual interface, users can adjust, remove, combine, and reuse AI proposals rather than treating them as finished edits. Thus, AI proposes \emph{what} may be useful to manipulate, semantic regions determine \emph{where} those manipulations are organized, and users retain control over \emph{how} they are ultimately configured.

We evaluate AI-composed controls in a within-subjects study with 12 intermediate-to-professional image-editing practitioners, comparing them with manually configured controls in the same editing environment. Participants strongly preferred working with AI-composed controls and described them as supporting creative exploration, while continuing to substantially refine AI proposals. At the same time, participants reported significantly lower felt control with AI composition, while comparative responses about control were more mixed. Interviews suggest that control depended not only on whether AI proposals could be modified, but also on whether participants understood why particular controls and values had been proposed. These findings suggest that making AI interpretations manipulable does not necessarily make them legible. Through this work, we contribute:

\begin{enumerate}[leftmargin=*]

\item Formative study findings on AI-assisted image-editing practices, showing how practitioners move between semantic goals and editing parameters, work with localized image regions, and seek opportunities for manual intervention when using AI.

\item \textbf{Atelier}, an image-editing system that translates semantic interpretations of an image and editing goal into persistent, context-specific parametric controls that users can inspect and revise.

\item Empirical findings showing how practitioners use AI-composed controls as starting points for exploration and refinement, while revealing a tension between the ability to manipulate AI proposals and the ability to understand the interpretations behind them.

\end{enumerate}






\section{Related Work}
\label{sec:related-work}

Our work builds on two related directions: research that helps users express high-level visual intentions and manipulate their realization in image editing, and research that externalizes AI interpretations as adaptive, persistent interaction artifacts. These areas motivate Atelier's approach of using image context and user intent to determine which familiar editing operations should become directly manipulable, where they should apply, and how they should be initialized.

\subsection{Semantic Intent and Parametric Image Editing}
\label{sec:rw-semantic-parametric}

Generative image editing increasingly allows users to describe desired changes at a semantic level instead of specifying individual editing operations. Recent systems combine language with sketches, masks, spatial annotations, and reference images to give users more direct ways to express both what should change and where that change should occur. MagicQuill~\cite{liu_magicquill_2025}, for example, combines coarse spatial interaction with generative editing, while Block and Detail~\cite{sarukkai_block_2024} and SketchFlex~\cite{lin_sketchflex_2025} support editing at different spatial and semantic granularities. Canvas-to-Image~\cite{dalva_canvas--image_2025} combines text, spatial layouts, and references to specify different aspects of an intended result, and PromptCharm~\cite{wang_promptcharm_2024} makes relationships between textual concepts and image regions inspectable and adjustable. Other work addresses the ambiguity of high-level intentions: Specify-and-Edit~\cite{iakovleva_specify_2024} decomposes underspecified instructions into more concrete edits, whereas Creativity-VLA~\cite{shen_empowering_2024} presents multiple possibilities that users can explore instead of committing to a single interpretation. These systems reduce the need for users to formulate a complete low-level editing procedure in advance.

A related line of work turns semantic properties themselves into objects or dimensions that can be manipulated. GANSpace~\cite{harkonen_ganspace_2020}, StyleCLIP~\cite{patashnik_styleclip_2021}, and Concept Sliders~\cite{gandikota_concept_2024} expose interpretable properties of generative models as continuous directions or controls. Interaction research has examined how such dimensions can be presented to users: GANSlider~\cite{dang_ganslider_2022} demonstrates both the opportunities and interaction costs of exposing multiple semantic dimensions, while Tweeq~\cite{hashimoto_tweeq_2025} focuses on compact and precise parameter manipulation for creative professionals. More recent systems move beyond a fixed set of semantic dimensions. DesignWeaver~\cite{10.1145/3706598.3714211} structures text-to-image exploration around design-relevant dimensions, while Brickify~\cite{10.1145/3706598.3714087} turns visual properties into reusable design tokens that can be directly manipulated. 


Across these works, some systems help users specify high-level intentions, while others expose semantic properties as interactive controls. Less explored is how context-dependent intentions can be translated into the familiar parametric operations of conventional image editing. This gap highlights an opportunity to bridge semantic intent and direct manipulation without requiring users to reconstruct that mapping themselves.

\subsection{Reifying AI Interpretations as Interactive Artifacts}
\label{sec:rw-reified-interaction}

The idea that interfaces can adapt to users, tasks, or context predates modern generative AI. Mixed-initiative interaction emphasizes that automated assistance should remain observable, correctable, and under appropriate user control~\cite{horvitz_principles_1999}. Systems such as SUPPLE~\cite{gajos_supple_2010} demonstrated that interfaces could be generated from models of users, devices, and tasks, while research on adaptive interfaces identified a recurring tension between relevance and predictability: changing which functionality is exposed can make an interface more useful in context while also making its behavior harder to anticipate~\cite{findlater_design_2009}.

Recent work uses generative models not only to produce content but also to construct interaction mechanisms. DynaVis~\cite{vaithilingam_dynavis_2024} translates natural-language visualization edits into persistent graphical widgets that can subsequently be manipulated. Bespoke~ \cite{nandy_bespoke_2024} uses model reasoning to compose task-specific interfaces around interpreted user intent, while generative and malleable interfaces allow interaction structures themselves to evolve with users' tasks~\cite{cao2025malleable,chen_generative_2025}. These approaches shift part of the model's contribution from generating a result toward determining how users can subsequently interact with it.

Several systems further emphasize \emph{reification}: turning otherwise transient prompts, interpretations, or configurations into persistent interactive objects. \emph{Cells, Generators, and Lenses}~ \cite{10.1145/3586183.3606833} introduces reusable objects for specifying inputs, combining model configurations, and inspecting outputs, allowing users to work with LLM interactions without reconstructing them through repeated conversation. AI-Instruments~\cite{10.1145/3706598.3714259} similarly embodies prompted graphical commands as reusable instruments that users can inspect, apply, and refine. PromptCanvas~\cite{amin_promptcanvas_2025} organizes parameterized prompt widgets on a persistent canvas. Across these systems, externalizing intermediate AI representations creates opportunities for direct manipulation, reuse, and reflection that are difficult to support when the interaction remains confined to a transient prompt--response exchange.

This perspective also connects to work on interaction substrates. Mackay and Beaudouin-Lafon~\cite{mackay2025substrates} describe substrates as persistent places that organize information and the operations that act upon it. Related work uses compositional structures as substrates for organizing human--AI co-creation~\cite{10.1145/3706598.3713401}. DesignTrace~ \cite{peng_designtrace_2026} applies this idea to visual design by reifying design rationale into interconnected semantic and visual artifacts that can be transformed, revisited, and reused across an evolving design process. Its semantic substrate supports extracting, creating, and composing design attributes, while its visual substrate supports global and localized visual editing.


These works show the value of making AI-mediated interactions persistent, reusable, and directly manipulable. At the same time, adaptive and generated interfaces introduce a tension between contextual relevance and predictability: the more a system determines what interaction elements should appear, the more important it becomes that users can understand, revise, and continue working with those choices. This gap highlights an opportunity for interfaces that make AI interpretations persistent parts of the interaction space without separating them from the familiar structures and operations through which users already work.


\section{Formative Study}
\label{sec:formative-study}

To understand how professional editors currently approach AI-assisted image editing and identify opportunities for improved interaction, we conducted a formative study with eight participants (F1-F8). 
Participants had 
experience with AI-powered editing tools such as Adobe Photoshop, Lightroom, Figma, and generative AI features. Through a semi-structured interview, we explored their editing workflows, use of AI-assisted tools, strategies for selecting and manipulating image regions, challenges encountered during editing, preferences for interaction techniques, and expectations for future AI-assisted interfaces (\autoref{tab:f-demographics}). Interview responses were analyzed using deductive thematic analysis by two researchers~\cite{braun_using_2006, braun_reflecting_2019,braun_one_2021}. First the two researchers familiarized themselves with the data, and independently coded the interviews aligned with the preliminary themes. The preliminary themes included: \textit{goal-orientated editing, mixed initiative interaction, context-aware editing, transparent AI assistance, and reflection.} Then the two researchers discussed, refined, or eliminated interpretations until consensus was reached.

\begin{table*}[htbp]
\centering
\caption{Formative study participants’ backgrounds and experience. Profession,
years of editing experience, age, gender, regularly used tools, and primary
editing work. All eight answered every open question.}
\Description{Backgrounds and image-editing experience of the eight formative
study participants. Participants span a range of professions, including
architecture, visual design, photography, software development, and other
creative work, with two to seven years of editing experience and ages from
24 to 40. They regularly use tools including Photoshop, Lightroom, GIMP,
Canva, and other image-editing software, and their primary work includes
color grading, compositing, retouching, graphic design, and illustration.}
\label{tab:f-demographics}

\footnotesize
\renewcommand{\arraystretch}{0.95}
\setlength{\tabcolsep}{4pt}

\begin{tabular}{@{}
    c
    l
    c
    c
    c
    p{4.2cm}
    p{5.0cm}
@{}}
\toprule
\textbf{ID} &
\textbf{Profession} &
\textbf{Yrs} &
\textbf{Age} &
\textbf{Gender} &
\textbf{Tools} &
\textbf{Primary work} \\
\midrule

F1 & Architect & 4 & 27 & m &
Photoshop, Lightroom, Affinity Photo &
Compositing, color grading \\

F2 & Videojournalist & 2 & 26 & m &
Photoshop, Lightroom &
Compositing, color grading \\

F3 & Hobbyist & 6 & 24 & m &
GIMP, DaVinci Resolve, Canva &
Color grading, graphic design \\

F4 & Tattoo artist & 7 & 27 & m &
Photoshop, Procreate &
Graphic design, illustration \\

F5 & Software dev (DevOps) & 5 & 40 & m &
Photoshop, Lightroom, GIMP &
Retouching, color grading \\

F6 & Visual designer & 6 & 28 & m &
Photoshop, Canva, AI tools &
Retouching, compositing, color grading, graphic design \\

F7 & Photographer & 3 & 34 & f &
Lightroom, Canva, Snapseed &
Color grading \\

F8 & UX designer & 5 & 26 & m &
Photoshop, Lightroom, AI tools &
Retouching, compositing, color grading, graphic design \\

\bottomrule
\end{tabular}
\end{table*}

\subsection{Findings}
\label{subsec:prestudy-themes}
Our analysis revealed four recurring themes that characterize how participants approach AI-assisted image editing and what they expect from future editing interfaces. Across participants, editing was described as a process that begins with semantic and perceptual goals, selectively incorporates AI to reduce effort, and relies on context-appropriate controls while preserving user agency. Participants also emphasized the importance of being able to inspect and recover from AI-generated changes. We describe these four themes below.

\subsubsection{Users Think in Semantic Goals Rather Than Technical Parameters}



Participants often described their editing intentions in terms of the visual or perceptual outcome they wanted to achieve, rather than the technical parameters needed to produce it. 
F3 described thinking in terms of \textit{``warmer, darker, more dramatic''} and added, \textit{``I think in feelings.''} F6 described this distinction more explicitly: \textit{``I usually start with goals first 
[...] like `make this warmer' or `bring focus to the subject,' or `make the background less distracting.' That helps me define what I want the viewer to feel rather than getting stuck in numbers.''} Technical parameters were still important, but often entered later as a means of realizing these intentions. F6 described shifting from an initial goal into adjustments of brightness, contrast, saturation, curves, and color balance, characterizing the process as \textit{``a back-and-forth process.''} Thus, participants did not necessarily think exclusively in either semantic or technical terms. Instead, their workflows often began with a desired visual outcome and subsequently translated that outcome into concrete editing operations.


\subsubsection{Users Value AI Assistance but Want to Remain in Control}

Participants valued AI for reducing the effort required for time-consuming editing operations, but did not view it as a replacement for manual editing. Instead, they selectively moved between AI and manual tools depending on the task. F1, for example, described this division as \textit{``For fast editing I prefer AI, but for final presentation I prefer manual editing.''} 
At the same time, participants were cautious about losing control over the final result. F8 characterized AI as \textit{``a hit or miss and you lack control for the final result.''} This uncertainty meant that AI-generated results were often treated as intermediate rather than final outcomes, with participants returning to manual tools when greater precision was needed. For some participants, retaining control was not only a response to imperfect AI outputs, but also an important part of creative ownership. F3 (Hobbyist) explained, \textit{``I think that art and editing is something that is very personal, each person has its own taste, its own trained eye of what they like and what not [...] It's something personal that an AI can't take over.''}


\subsubsection{Users Want Context-Aware Editing Controls}

Participants emphasized that useful editing controls depend on both the image content and the particular attribute being manipulated. 
They envisioned interfaces that adapt their controls to the current editing context. 
F8, for example, explained, \textit{``I think every control is fine depending on the context. For skin color maybe a tuned down color picker would be nice. For night and day a slider, for something else a toggle might be fine.''}. Other participants described context-awareness in terms of the relationships between elements in an image. F4, for example, wanted assistance that could account for lighting relationships rather than treating individual elements independently, describing a tool that could identify \textit{``where the light source is coming from''} and use this information to match lighting and shadows within the composition. This suggests that context-aware assistance may need to reason not only about individual objects, but also about their visual relationships. Participants also envisioned AI helping identify the semantic content of editable regions. F6 wanted to \textit{``immediately understand and refine what the tool thinks that region is (skin, hair, sky, shadow, etc.), instead of just seeing a flat mask.''} They further suggested \textit{``Contextual controls based on selection would be very useful, especially if they are intelligent and minimal''} but only as an addition, never a replacement for the general controls practitioners already know (F3).


\subsubsection{Transparency and Recoverability Support Trust}

Participants emphasized that working effectively with AI required being able to inspect and evaluate its behavior, particularly because generated results were not always predictable. F8 characterized AI as \textit{``a hit or miss,''} while F5 described the difficulty of repeatedly adjusting prompts without reaching the intended result: \textit{``Sometimes you improve your prompt, but still you are missing the expected result.''}. These experiences made mechanisms for understanding and evaluating AI-generated changes particularly important. Before-and-after comparisons were one recurring strategy for inspecting changes. F3 explained, \textit{``I usually use the toggle functionality to see a before and after---when it comes to AI I think that transparency is key.''} F1 similarly emphasized comparison when evaluating edits, stating, \textit{``I want to see the before and after.''} Such mechanisms allowed participants to assess AI-generated changes against the original image rather than accepting an output at face value. Participants also wanted greater visibility into how AI-assisted effects were produced. F6 explained, \textit{``If I can't see how an effect is being applied, it becomes harder to trust the result or refine it properly.''} Here, transparency was not necessarily about understanding the internal workings of the model, but about making the relationship between an editing action and its visual consequence legible. Finally, the participants emphasized the importance of being able to recover from unsuccessful results. F8 described strategies such as \textit{``revert, save intermediate steps, and regenerate''} when an AI operation failed, while F4 described abandoning an AI-assisted approach when repeated outputs did not correspond to their intention. These responses indicate that participants did not expect AI to always be correct; instead, they wanted failures to remain inexpensive to inspect, reverse, and refine.

\subsection{Design Goals}
Based on these findings and prior work, we derived three design goals for AI-assisted image-editing interfaces.


\begin{enumerate}[leftmargin=*, itemsep=1mm, label=\textbf{DG\arabic*}]



\item \textbf{Bridge semantic intentions and parametric operations.}
Participants often began with high-level visual goals before considering the technical parameters needed to realize them. Interfaces should therefore translate semantic intentions into inspectable and editable parametric operations, without requiring users to identify the relevant controls themselves.

\item \textbf{Decompose editing into spatially scoped sub-tasks.}
Participants expected different image regions to benefit from context-appropriate controls while having access to familiar general-purpose tools. Interfaces should let users decompose an image into the regions they are already thinking about — an object, a subject, a sky — and scope controls to that region, whether those controls are configured manually or composed by an AI, while keeping the full general-purpose surface available at all times.

\item\textbf{Preserve user control and authorship.}
Participants emphasized that trust in AI develops incrementally as users inspect and evaluate its behavior during collaboration. Interfaces should therefore support comparison, previews, undo where appropriate, and localized refinement, as these mechanisms are essential to preserving users’ sense of control when AI composes the editing controls. 




\end{enumerate}

\section{Atelier}
\label{sec:interaction-concept}

Based on the design goals we build Atelier, which is a canvas-based image editor that augments a conventional image editing interface with AI-composed controls. Images, extracted regions, and editing widgets coexist on the canvas, while a standard control panel provides access to the full set of conventional editing operations. 
Users can invoke AI assistance as needed and refine the proposed controls through direct manipulation. We provide a supplementary video to this submission showing the key interactions supported by the system.



\subsection{Interface and Interaction}
\label{sec:interface-interaction}


\subsubsection{Interpret Image and Intent}
\label{sec:interpret}

\begin{figure*}
    \centering
    \includegraphics[width=\textwidth]{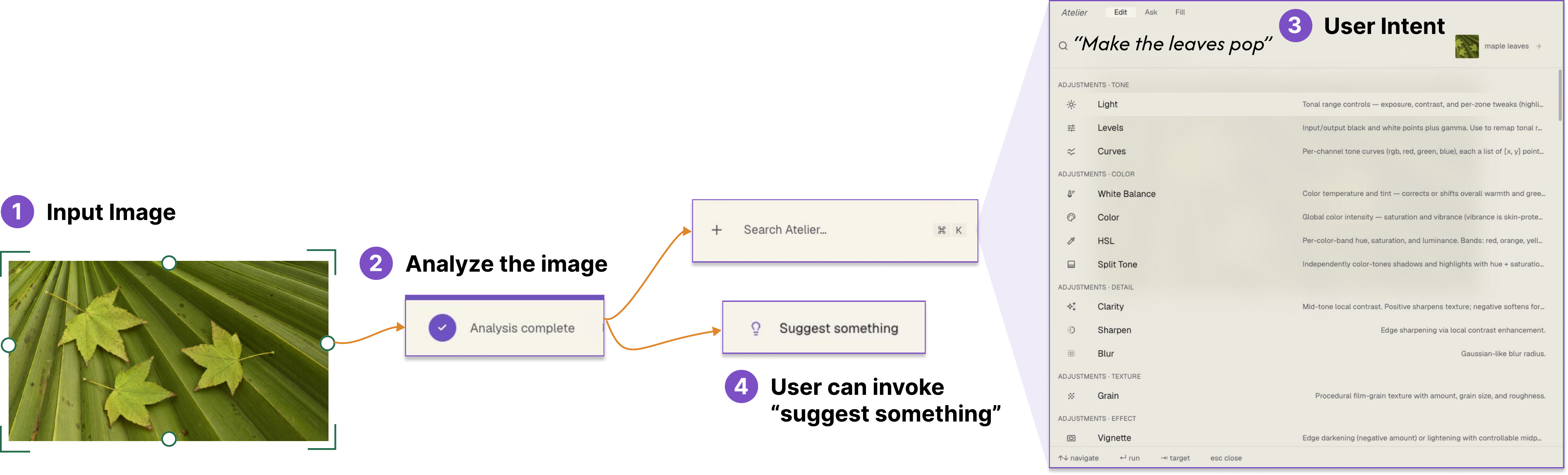}
    \caption{Interpreting image context and identifying relevant editing dimensions in Atelier. After an image is uploaded (1), Atelier analyzes its visual content and characteristics (2). The user can then provide an editing intention such as \textit{``make the leaves pop''} (3), which Atelier interprets in relation to the current image to identify relevant editing dimensions, such as leaf contrast. Alternatively, the user can invoke \emph{Suggest something}, allowing Atelier to propose a potentially useful editing direction from the image context alone (4), illustrated here with a detail-related adjustment.}
    \Description{Left-to-right diagram in four numbered steps, all converging on one command-palette panel. (1) An input image of green maple leaves, shown with selection handles. (2) An arrow leads to an ``Analysis complete'' indicator, labeled Analyze the image. From there, two arrows branch to two entry points that both open the same palette panel on the right: a ``Search Atelier'' field, and a ``Suggest something'' button, labeled (4) User can invoke ``suggest something.'' The palette panel on the right has three tabs, Edit, Ask, and Fill, and a search box containing the typed text ``Make the leaves pop,'' labeled (3) User Intent, with a small thumbnail and the label ``maple leaves'' beside it. Below the search box, a scrollable list of editing dimensions is grouped under five headers: Tone (including Light, Levels, and Curves), Color (including White Balance, Color, HSL, and Split Tone), Detail (including Clarity, Sharpen, and Blur), Texture (Grain), and Effect (Vignette). Each entry has a short one-line description beside it. A footer shows keyboard navigation hints.}
    \label{fig:interaction-interpret}
\end{figure*}



Once an image\footnote{Image generated using OpenAI’s GPT-5.6 Sol model through the ChatGPT interface (details in the Appendix)} is uploaded, Atelier analyzes its visual content and characteristics, including recognizable objects and regions, dominant colors, tonal properties, and other features that may provide context for editing. As illustrated in \autoref{fig:interaction-interpret}, this process begins with the current image (\autoref{fig:interaction-interpret}-1) and an analysis of its visual context (\autoref{fig:interaction-interpret}-2). Users can then express an editing intention in terms of the visual result they want to achieve rather than the technical operations required to produce it. For example, a user may request to \textit{``make the leaves pop''} (\autoref{fig:interaction-interpret}-3). Atelier interprets this intention together with the current image rather than treating the phrase as a fixed command. What it means for the leaves to ``pop'' depends on properties such as their current brightness, contrast, color, and visual relationship to their surroundings.

Atelier's context-aware interpretation therefore establishes the semantic starting point for \textbf{DG1}: users can begin with a desired visual effect without first identifying the technical parameters needed to achieve it. 

\subsubsection{Identify Relevant Editing Dimensions}
\label{sec:identify_dimensions}

Based on this interpretation, Atelier identifies a small set of \emph{editing dimensions} that may contribute to the intended effect. These dimensions describe perceptually meaningful aspects of the image that could be manipulated, such as lighting, contrast, color, or detail. 


Editing dimensions can also be identified without an explicitly stated intention. After analyzing the image, users can invoke \emph{Suggest something}, as shown by the lower branch in \autoref{fig:interaction-interpret}. Atelier then uses the image context itself to identify a potentially useful editing direction and proposes corresponding dimensions (\autoref{fig:interaction-interpret}-4). These dimensions support \textbf{DG1} by turning a high-level intention into concrete aspects of the image that can be edited. At this stage, Atelier determines \emph{what} may be useful to manipulate.

\begin{figure*}
    \centering
    \includegraphics[width=\textwidth]{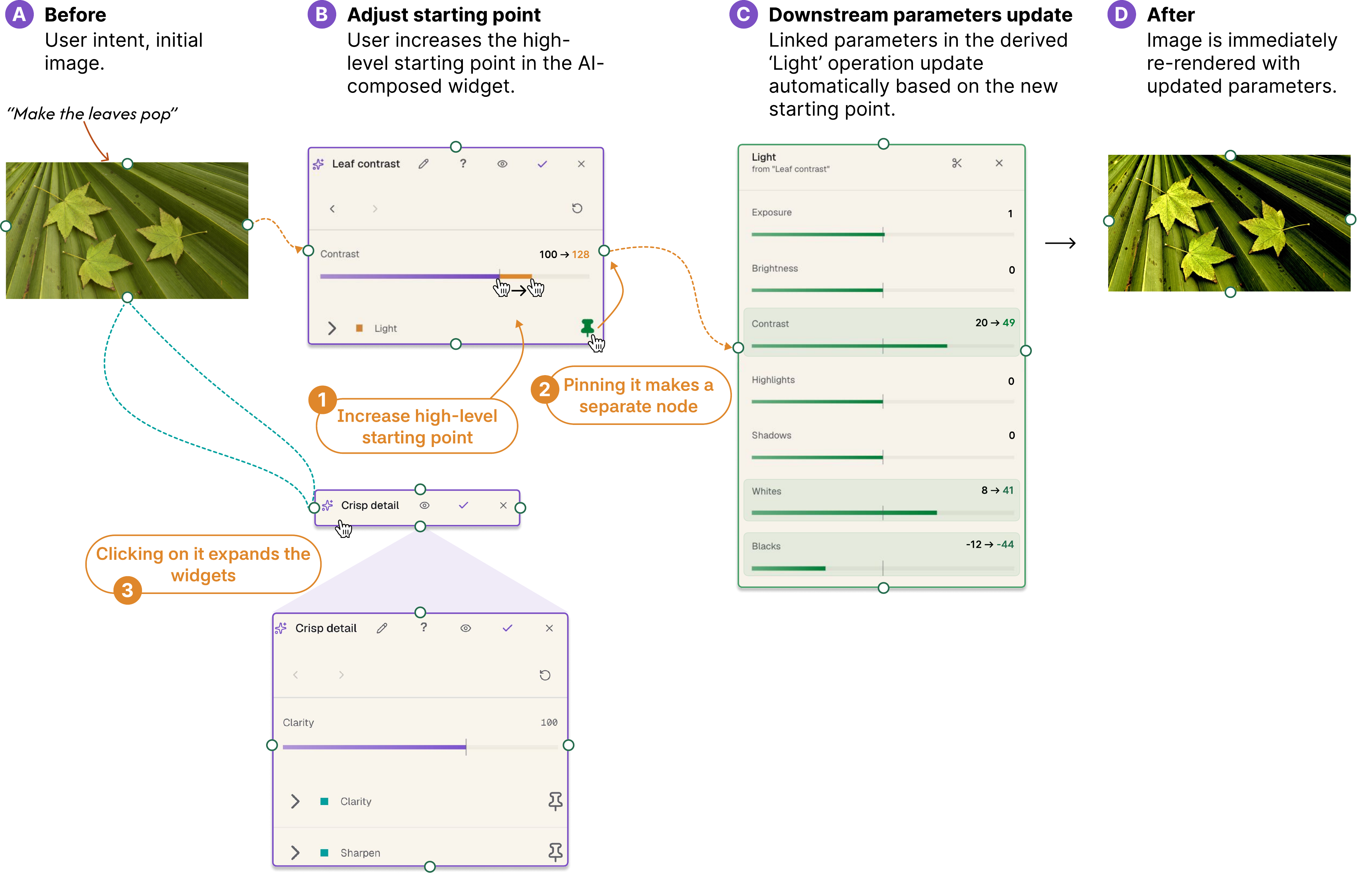}
    \caption{AI-composed widgets in Atelier. User gives an image and the intention ``Make the leaves pop,'' (A) Atelier composes two high-level controls (Leaf contrast and Crisp detail). Increasing its starting point (B) propagates the change to the more specific parameters of the underlying Light operation, such as contrast, whites, and blacks (C), and immediately re-renders the image (D). Users can also pin an operation to expose it as a separate persistent node and expand compact widgets to access their constituent controls.}
    \Description{Left-to-right interaction sequence in Atelier, using an image of three maple leaves, across four labeled steps. In A, the initial leaf image is shown with the typed intention ``Make the leaves pop.'' In B, two widgets are connected to the image by dashed lines: a Leaf contrast widget, where the user drags its high-level contrast value from 100 to 128, and a collapsed Crisp detail widget. Three callouts label the interaction: increasing the high-level starting point; pinning the widget's underlying Light operation to expose it as a separate node; and clicking the Crisp detail widget to expand it, shown below opening into Clarity and Sharpen controls. In C, the expanded Light operation shows the high-level change propagating to its detailed parameters: contrast changes from 20 to 49, whites from 8 to 41, and blacks from -12 to -44, while exposure, brightness, highlights, and shadows stay unchanged. In D, the resulting image has visibly stronger contrast and more prominent leaves.}
    \label{fig:compose_widgets}
\end{figure*}

\subsubsection{Compose Parametric Widgets}
\label{sec:parametric-widgets}





Atelier translates the identified editing dimensions into concrete parametric operations. A single stated intention can produce a composition of one to six independently refinable controls. \autoref{fig:compose_widgets} illustrates this translation for the intention \textit{``make the leaves pop.''} Atelier composes two operations (\textit{Leaf contrast} and \textit{Crisp detail}) that together realize the identified editing dimensions (\autoref{fig:compose_widgets}). A planner selects which operations from Atelier's predefined registry should participate in the composition, while a resolver assigns their initial parameter values. Because the operations come from this registry, their parameter types, valid ranges, and corresponding interface controls are already defined.

Instead of applying the system's interpretation as a finished transformation, Atelier externalizes it as familiar editing operations and proposed parameter values that users can inspect and manipulate \textbf{(DG1)}. 
AI composition does not replace Atelier's conventional editing interface. A dedicated sidebar provides persistent access to the complete set of standard editing operations, allowing users to manually choose and adjust controls independently of the semantic composition process. AI-composed widgets and sidebar controls draw from the same operation registry and modify the same underlying editing state. A change made through an AI-composed widget is reflected in the corresponding conventional control, and vice versa.

\begin{figure*}[htbp]
\centering
\includegraphics[width=0.7\textwidth]{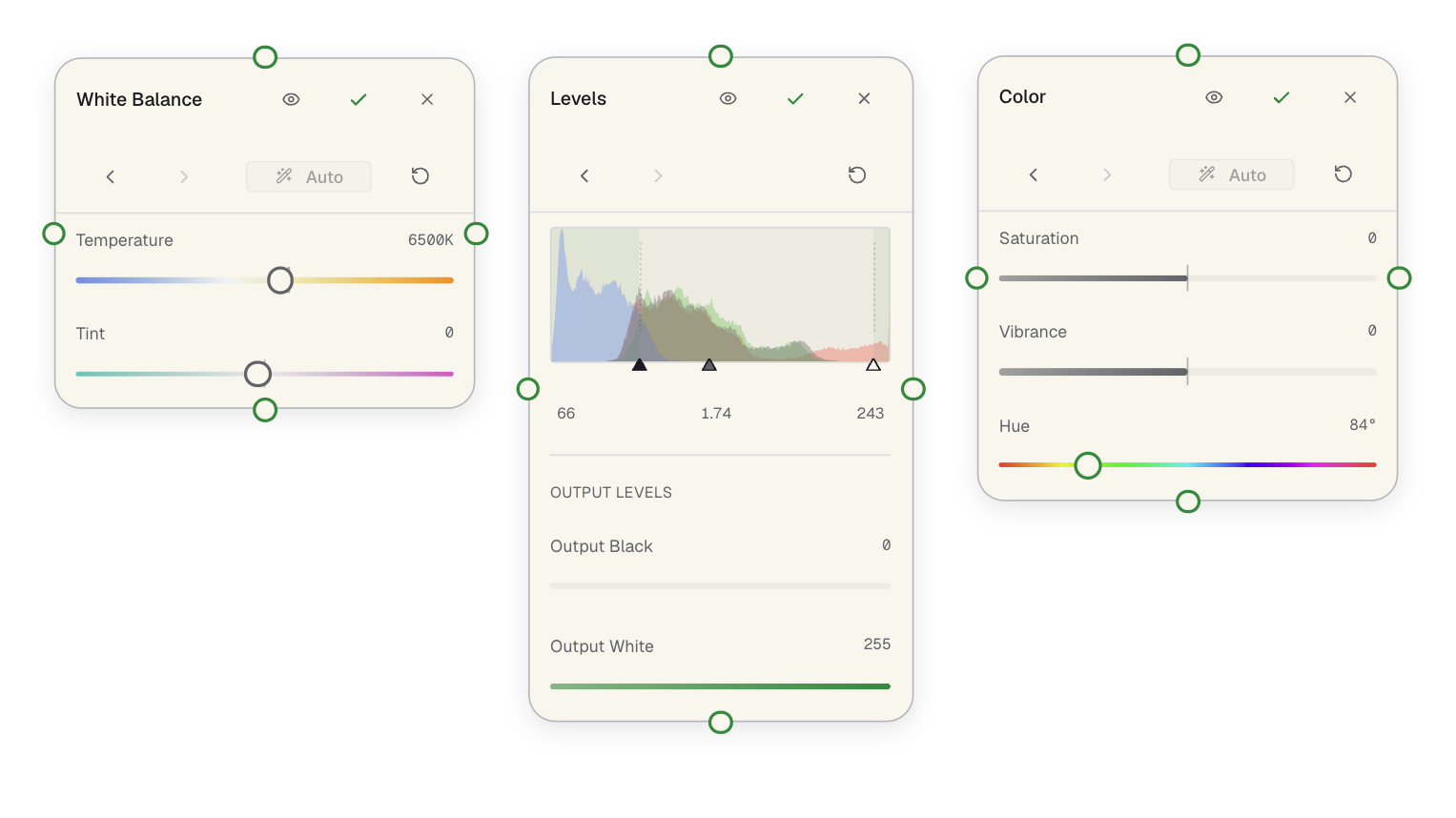}
\caption[Anatomy of a widget.]{Three widgets sharing one shell: a header (name, visibility toggle, accept and dismiss), a row for stepping between proposals, \emph{Auto}, and one-click reset, and a parametric body that varies by operation (white balance, \emph{Levels}, \emph{Color}). The green handles are the canvas ports that connect the widget to its image-node.}
\Description{Three widget panels shown side by side, sharing the same header and toolbar row, with different controls below. White Balance shows a Temperature slider spanning blue to orange, set to 6500K, and a Tint slider spanning green to magenta, set to 0; its toolbar includes an Auto button. Levels shows a histogram with three overlapping color-channel distributions, three triangular point markers below it labeled 66, 1.74, and 243, and an Output Levels section with an Output Black slider set to 0 and an Output White slider set to 255; its toolbar has no Auto button. Color shows Saturation and Vibrance sliders both set to 0, and a Hue slider spanning the full color spectrum, set to 84 degrees; its toolbar includes an Auto button.}
\label{fig:widget-anatomy}
\end{figure*}

All operations use the same widget vocabulary. As shown in \autoref{fig:widget-anatomy}, widgets share a common shell while their parametric body adapts to the represented operation. The examples show White Balance, Levels, and Color, whose bodies expose different controls while retaining common interaction elements. 
Canvas ports additionally keep widgets visibly connected to the image or region they affect. Instead of requiring users to adjust every parameter independently from the outset, Atelier presents each AI composition through a shared high-level driver named after its intended effect. In \autoref{fig:compose_widgets}-B, for example, the \textit{Leaf contrast} driver is increased from 100 to 128. This adjustment propagates across the underlying parameters that jointly realize the effect. Expanding the associated \textit{Light} operation reveals these changes directly (\autoref{fig:compose_widgets}-C): contrast increases from 20 to 49, whites from 8 to 41, and blacks from $-12$ to $-44$, while parameters unrelated to the adjustment remain unchanged. Users can move from this coordinated adjustment to finer-grained control without leaving the composition. Compact operations can be expanded to expose their constituent parameters, as shown for \textit{Crisp detail}, and an operation such as \textit{Light} can be pinned to the canvas as a separate persistent widget. Users may then manipulate individual parameters directly or continue working with the operation independently of the original composition.

Atelier thus provides \textbf{parametric granularity}. Users can move between a high-level representation of an intended effect and the lower-level parameters that realize that effect. The ability to expand, separate, reset, dismiss, and manually refine AI-composed operations also implements mechanisms intended to support \textbf{DG3}. 

\begin{figure*}
    \centering
    \includegraphics[width=\textwidth]{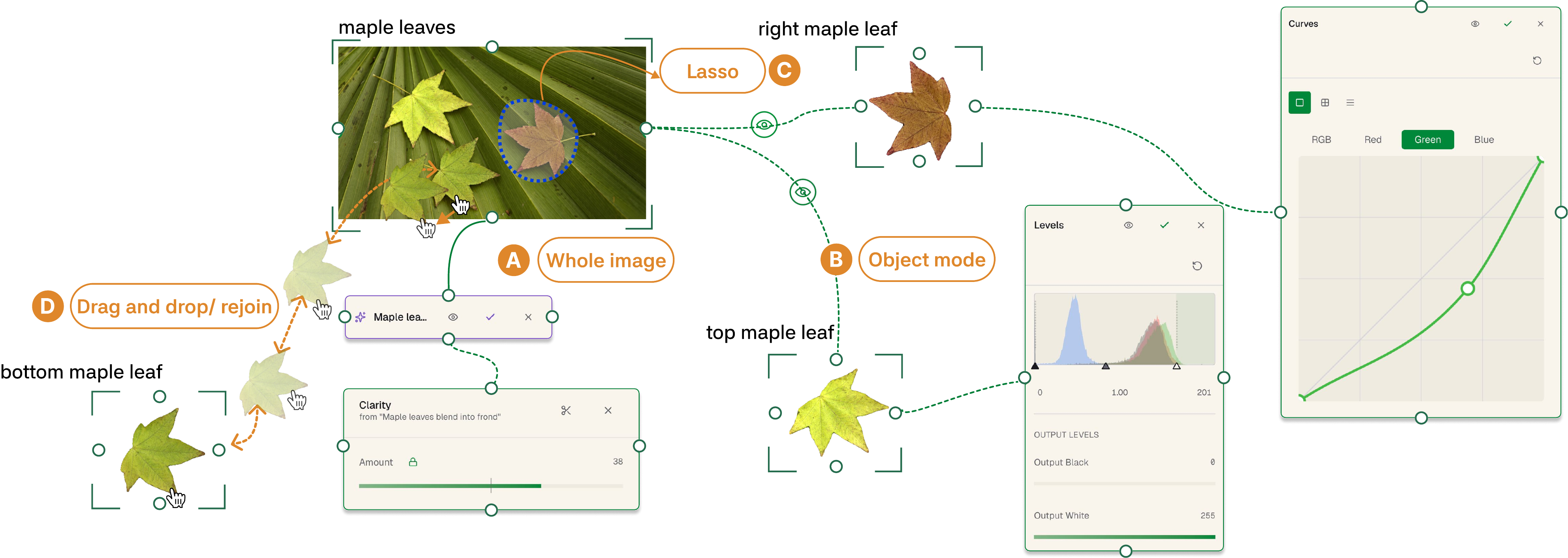}
    \caption{Establishing spatial targets in Atelier. An edit can apply to the whole image (A), to an object selected in \emph{Object Mode} (B), or to an arbitrary region defined with a lasso (C). Once established, these targets persist as addressable regions that can be connected to subsequent editing operations. In the \textit{Object Mode}, after clicking on the objects they can be dragged and dropped anywhere on the canvas and dragged regions can be rejoined with the source image after editing (D).}
    \Description{Diagram on an Atelier canvas showing four ways to define an edit's spatial target, illustrated with maple leaves. In A, the whole image is connected by a dashed line to a collapsed widget outlined in purple to mark it as AI-composed, labeled ``Maple lea...''; this widget connects to a Clarity panel below it, labeled ``from Maple leaves blend into frond,'' with its Amount value set to 38. In B, one leaf has been extracted as separate objects in Object Mode, labeled top maple leaf. It is connected by a dashed line to its own widget, a Levels panel,. In C, a dotted lasso outline is drawn around a third leaf within the image, with an arrow pointing to a Lasso label, illustrating the lasso-selection action. It leads to a separate object titled right maple leaf, and is connected to a Curves panel. In D, a fourth leaf is shown being clicked and dragged along a curved path to a new position outside the original image, labeled bottom maple leaf, illustrating that Object Mode selections can be dragged anywhere on the canvas and dragged regions can be rejoined with the source image after editing.}
    \label{fig:spatial-targets}
\end{figure*}

\subsubsection{Identify and Scope Regions}
\label{sec:regions}

Alongside determining \emph{what} may be useful to manipulate, Atelier determines \emph{where} an edit should apply. As illustrated in \autoref{fig:spatial-targets}, the spatial target may encompass the complete image when an adjustment is global (\autoref{fig:spatial-targets}-A), correspond to a recognizable object selected in Object Mode (\autoref{fig:spatial-targets}-B), or describe an arbitrary region defined manually with the lasso tool (\autoref{fig:spatial-targets}-C). An adjustment initially proposed for the complete image can instead be redirected to a particular object, or an object-level target can be further narrowed using a lasso or mask.


Once established, a target remains available on the canvas beyond the operation that created it. Objects selected in Object Mode or via Lasso can be dragged out of the image and positioned independently on the canvas (\autoref{fig:spatial-targets}-D). Dragging the object back onto its source image rejoins it at its original position while preserving the edits made to the extracted object. The extracted content is reintegrated into the source as a separate layer. In line with \textbf{DG2}, these interactions allow users to decompose an image into regions corresponding to the sub-tasks they are already considering and bind controls to the specific target for which an adjustment is relevant. In this way, Atelier provides \textbf{spatial granularity}.

\subsubsection{Revisit, Compare, and Reuse}
\label{sec:revisit}

\begin{figure*}
    \centering
    \includegraphics[width=\textwidth]{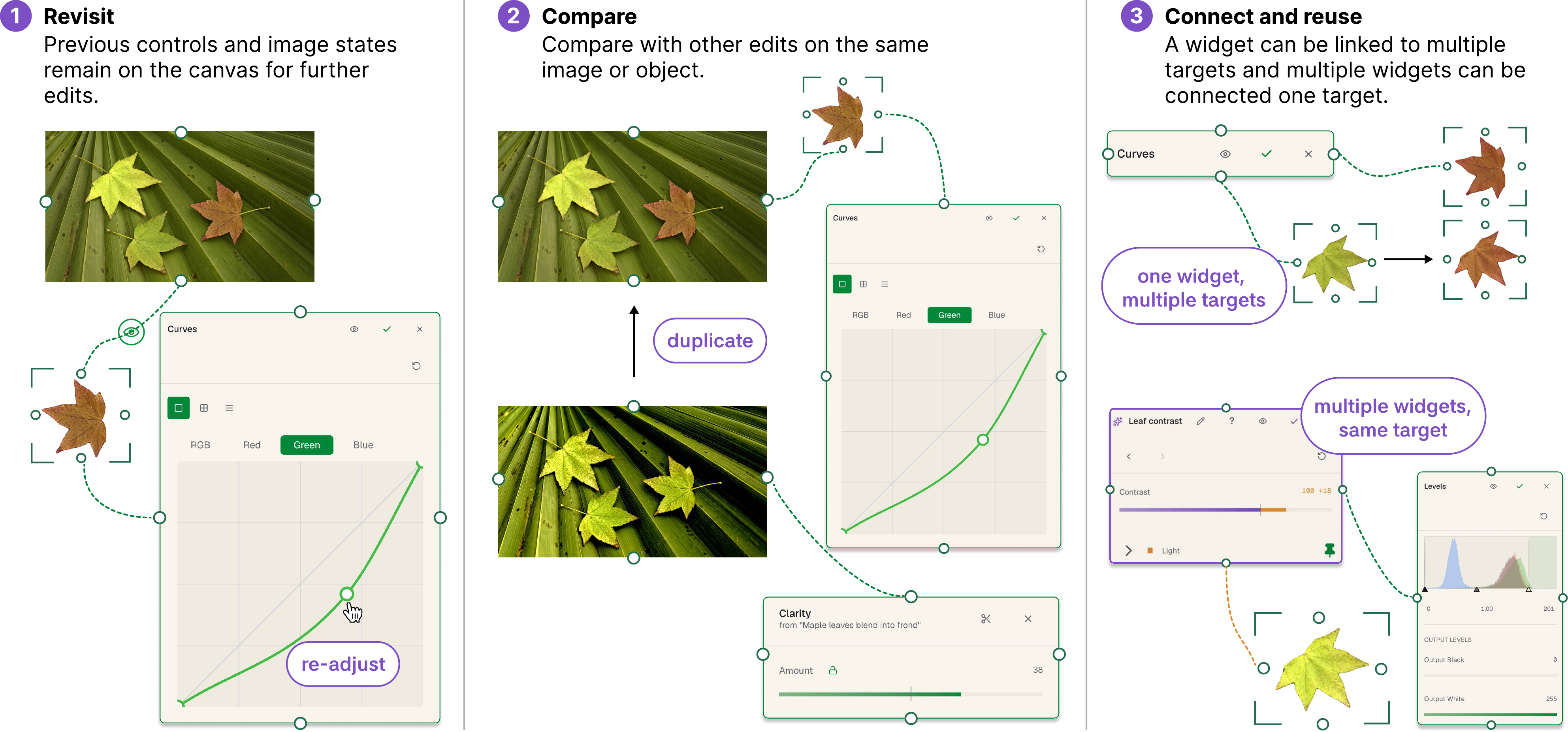}
    \caption{Revisiting, comparing, and reusing edits in Atelier. Previous controls and image states remain available on the canvas and can be readjusted as the editing process develops (1). Users can duplicate an image or object to explore and compare an alternative edit independently (2). Widgets and visual targets can also be reconnected, allowing one editing configuration to be reused across multiple targets or several widgets to contribute to the same target (3).}
    \Description{A three-part Atelier interaction diagram. In panel 1, labeled Revisit, an extracted leaf remains connected to a Curves widget on the canvas, where the user drags a point on the curve, next to a ``re-adjust'' callout. In panel 2, labeled Compare, the original image and an extracted leaf are connected to a Curves widget; a ``duplicate'' callout leads to a second, independently lit copy of the image below, which is connected instead to a Clarity panel labeled ``from Maple leaves blend into frond,'' with its Amount set to 38. In panel 3, labeled Connect and reuse, a Curves widget is connected to two separate leaf objects, next to a callout reading ``one widget, multiple targets.'' Below it, a Leaf contrast widget, with its Contrast value set from 100 to 128, and a Levels widget are both connected to the same leaf object, next to a callout reading ``multiple widgets, same target.''}
    \label{fig:revisit}
\end{figure*}

Once created, Atelier's controls and visual elements remain on the canvas. This persistence allows users to return to earlier decisions, compare alternative directions, and recombine existing controls and targets as their editing process develops (\autoref{fig:revisit}).

\textbf{\textit{Revisit.}} Previous controls remain directly manipulable, allowing users to return to an earlier edit and continue adjusting it. As shown in \autoref{fig:revisit}-1, a previously created \textit{Curves} control remains connected to its target and can be readjusted as the desired appearance changes. Users can similarly expand composed controls, modify individual parameters, reset them, or remove proposals that no longer fit their intention.

\textbf{\textit{Compare.}} Users can duplicate an image or extracted object and continue editing the copy independently, allowing alternative image states to coexist on the canvas. \autoref{fig:revisit}-2 illustrates this by showing two versions of the same image with different adjustments. Rather than committing immediately to one result, users can keep these alternatives visible and compare how different controls or parameter configurations affect the image. Earlier states can therefore become starting points for further exploration. 

\textbf{\textit{Connect and reuse.}} Controls and visual targets are linked explicitly on the canvas and can be reconnected without recreating the underlying edit. As shown in \autoref{fig:revisit}-3, a single widget can be connected to multiple targets, allowing the same editing configuration to be reused across all. Conversely, multiple widgets can be connected to the same target, allowing independently created adjustments to contribute to a shared result. Users can reuse an existing edit by redirecting or extending its connections instead of specifying the same operations again.

These interactions keep editing persistent and revisable: AI-composed operations and their spatial targets remain visible, while unsuccessful directions can be adjusted, removed, or compared \textbf{(DG3)}. This persistence can also support sensemaking and intentmaking~\cite{suhSensecapeEnablingMultilevel2023,intentmaking}, as users revisit prior edits, compare alternatives, and refine both their understanding of the current state and what they want to achieve.

\subsection{Technical Implementation}
\label{sec:technical-implementation}

\begin{figure*}
    \centering
    \includegraphics[width=0.85\textwidth]{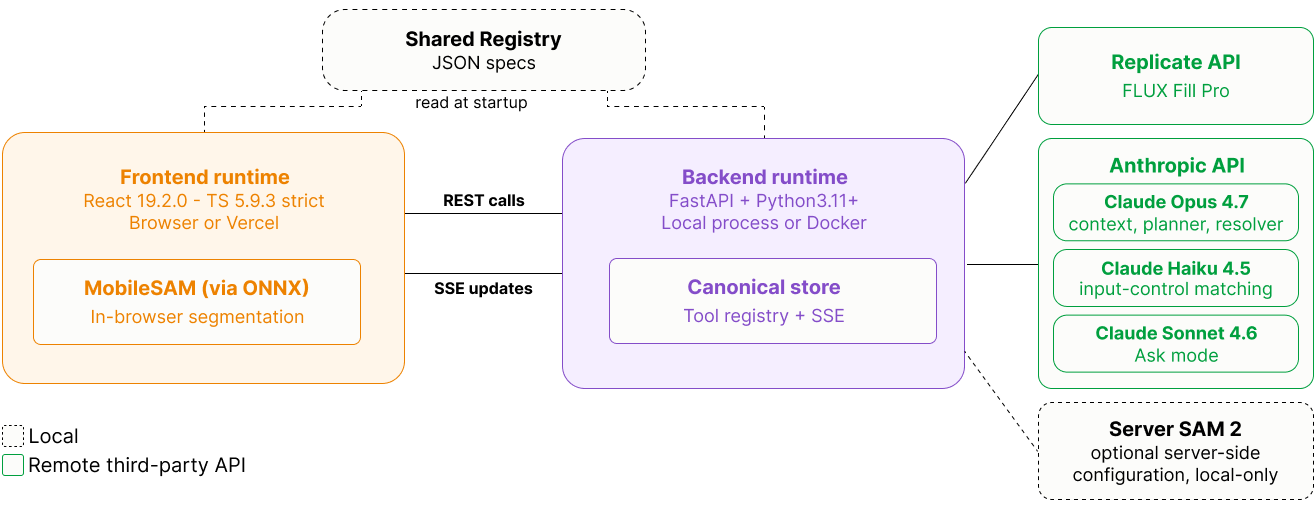}
    \caption{The system architecture of Atelier. The frontend is a React application in TypeScript, that renders the canvas, node workspace, and editing controls, communication with the backend over REST calls. The backend is a FastAPI application, that manages the parameter state and stores outputs produced by vision-language models. It communicates changes to the frontend over a SSE stream. The shared registry is a set of JSON files both runtimes read straight off the disk at startup. The backend calls the Anthropic API for the visual-language model and the Replicate API for the Generative Fill with FLUX Fill Pro. The frontend uses MobileSAM through ONNX Runtime in the browser for object selection, with the possibility of server-side configuration using SAM2.}
    \Description{Block diagram of Atelier's architecture, with a legend for local (dashed border) versus remote (green border) components, and six components: Frontend runtime, backend runtime, shared registry, Replicate API, Anthropic API, Server SAM 2. Two main blocks sit at the center: Frontend runtime, in orange, on the left, and Backend runtime, in purple, on the right, connected by two solid lines labeled REST calls and SSE updates. Frontend runtime is implemented in React 19.2.0 and TypeScript 5.9.3 strict, running in the browser or on Vercel, and contains a nested box for MobileSAM (via ONNX), used for in-browser segmentation. Backend runtime is implemented in FastAPI and Python 3.11+, running as a local process or in Docker, and contains a nested box for the canonical store, holding the tool registry and SSE. Above both blocks, a dashed Shared Registry box, labeled JSON specs, connects to each by a dashed line labeled read at startup, marking it as local. To the right of the backend, green-bordered boxes mark remote third-party APIs: Replicate API, for FLUX Fill Pro, and Anthropic API, which contains three nested model boxes: Claude Opus 4.7 for context, planner, and resolver; Claude Haiku 4.5 for input-control matching; and Claude Sonnet 4.6 for Ask mode. Below these, a dashed Server SAM 2 box, connected to the backend by a dashed line, marks an optional, local-only server-side configuration.}
    \label{fig:architecture}
\end{figure*}

Atelier uses a decoupled frontend--backend architecture (see \autoref{fig:architecture}). The React frontend renders the canvas, node workspace, and editing controls, while a Python backend built with FastAPI manages the parameter state that determines the rendered image and stores outputs produced by the vision-language models. 

\subsubsection{Image Analysis and Operation Planning}
\label{sec:image-analysis}

When an image is loaded, Claude Opus~4.7 produces a structured image context describing properties such as salient regions, dominant colors, tonal characteristics, and potential editing issues. This context is cached and reused by subsequent model calls. Given this image context, the user's intention, and any selected target region, a Claude Opus~4.7~\footnote{https://www.anthropic.com/news/claude-opus-4-7} \emph{planner} selects relevant operations from Atelier's predefined tool registry. The tool registry specifies the available operations, their parameters and valid ranges, and their associated control types. The planner determines \emph{which} operations should be exposed and grouped together, while their concrete parameter values are resolved in the subsequent composition stage. The same mechanism supports autonomous suggestions. When the user requests a suggestion without specifying an edit, the planner uses the cached image context to identify a potentially useful editing direction. For corrective suggestions tied to measurable image properties, Atelier additionally checks the relevant condition on a downsampled rendering before presenting the proposal.


\subsubsection{Widget Composition and Value Resolution}
\label{sec:widget-composition}

After the planner selects relevant operations, a second Claude Opus~4.7 call acts as the \emph{resolver}, assigning their initial parameter values. Related parameters are resolved jointly to form a coherent editing proposal. Each parameter is then rendered using the control type defined in the editing registry---such as a slider, curve, color control, or toggle---and bound to the same shared editing state as the conventional controls. The same pipeline is used whether a widget originates from a toolrail action, command-palette request, or autonomous suggestion. For the command palette, Claude Haiku~4.5~\footnote{https://www.anthropic.com/news/claude-haiku-4-5} performs latency-sensitive matching between natural-language input and available commands, while Claude Sonnet~4.6~\footnote{https://www.anthropic.com/news/claude-sonnet-4-6} supports \emph{Ask} queries about the current image without modifying the editing state. For widgets containing multiple parameters, Atelier additionally creates a higher-level \emph{driver} control that interpolates between the image state before the proposal and the resolver-generated values. Individual parameters remain directly adjustable; once a user modifies one manually, it is excluded from subsequent driver updates to preserve the refinement. Generated widgets remain reversible. Applying a widget incorporates its current values into the edit stack, while dismissing it restores the affected parameters to their previous state. Individual parameters and the complete editing session can likewise be reset.


\subsubsection{Object-Scoped Controls and Provenance}
\label{sec:object-controls}

Atelier represents spatial editing targets as Objects. For object selection, Atelier uses MobileSAM~\footnote{https://github.com/chaoningzhang/mobilesam} through ONNX Runtime~\footnote{https://onnxruntime.ai} directly in the browser, with an optional server-side configuration using SAM~2~\cite{ravi_sam_2025}. When the user clicks an object, the segmentation model converts the click into a pixel mask, which Atelier stores as an Object. Spatial selection is handled independently from language-model reasoning: the segmentation model determines \emph{where} an edit applies, while the Claude-based planner and resolver determine \emph{what} operations and parameter values should be applied. Objects may also originate from manually created selections or regions introduced through an AI-generated proposal. 
Atelier provides two lasso-based region-selection methods. The \emph{ lasso} uses a custom, dependency-free implementation that simplifies the stroke while drawing, rejects accidental clicks based on area, and fills the closed outline with a scanline algorithm to create the mask. The \emph{magic lasso} uses the rough outline as a spatial hint for MobileSAM to predict a more precise pixel-level mask, falling back to the original lasso shape if the prediction is implausible. Both selections are stored in the same \textit{Object} representation. 

\subsubsection{Generative Fill}
\label{sec:generative-fill}

Atelier uses FLUX Fill Pro through Replicate\footnote{https://replicate.com/black-forest-labs/flux-fill-pro} for edits that require generating new pixels. Users select a region and describe the desired content, after which FLUX performs inpainting and returns raster content that Atelier places on a new layer. Unlike the planner--resolver pipeline for parametric adjustments such as exposure or white balance, generative fill synthesizes new visual content. Because similar functionality is available in conventional editors such as Photoshop, we treated it as a baseline capability and made it available in both study conditions. 

\section{Study Design}



Our evaluation focuses on using image context and editing intentions to compose manipulable parametric controls. We therefore designed a within-subjects, mixed-methods study comparing Atelier's AI-composed controls with a Baseline that retained the same canvas, conventional editing controls, and commonly available AI-assisted capabilities, including object segmentation, lasso and generative fill. By holding these capabilities constant and varying the semantic and parametric control-composition layer, we aimed to examine how AI involvement in selecting and initializing familiar editing operations changes practitioners' editing experience, interaction behavior, and sense of control. We conducted the study with 12 image-editing practitioners. The study was reviewed and approved by the institution's Research Ethics Committee. Within this comparison, we investigated the following research questions:

\begin{enumerate}[label=\textbf{RQ\arabic*}]

    \item How can image-editing interfaces translate high-level editing intentions into context-specific, directly manipulable parametric controls?

    \item How does translating editing intentions into AI-composed parametric controls affect practitioners’ creative exploration and editing experience compared with manually configuring controls?

    \item How does AI involvement in selecting and initializing parametric controls affect practitioners’ felt control and sense of authorship when its proposals remain adjustable, reversible, and optional?
\end{enumerate}










\subsection{Experimental Setup}
\paragraph{Participants}

Our target users are intermediate to professional image-editing practitioners familiar with conventional tools and workflows. We focus on users with some experience of AI-assisted editing, without assuming expertise in prompting. We recruited twelve participants, evenly split between intermediate and professional experience based on self-assessment. 
Nine participants identified as male and three as female; mean age was 28.4 years ($SD=2.2$). Participants were recruited through convenience sampling (details provided in Appendix~\ref{app:participants}). 

\paragraph{Conditions}
We chose two conditions: AI-composed and the baseline condition, that had the same interface, including the canvas, the side panel with the photo editing controls, and the command palette. The AI-composed condition added autonomous suggestions (\ref{sec:identify_dimensions}), and composed parametric widgets (\ref{sec:parametric-widgets}).
The baseline condition did not include the context-aware adaptive widget, but did include the AI-backed capabilities click-to-segment selection (\ref{sec:object-controls}), generative fill (\ref{sec:generative-fill}), and the palette's Ask mode (\ref{sec:widget-composition}). This allowed us to study how this changes the experience of the context aware, granular editing, while keeping the baseline comparable to common photo editing applications.

\paragraph{Photo-Editing Briefs}
\label{sec:tasks}
Participants were asked to edit two images (in Appendix \ref{app:ms-stimuli}) and received a three-part brief.
The first part of the brief was a corrective task, asking to restore the degraded photo so the scene looks natural and true to life, each brief naming its image's defect. The target image was given, to allow the editing steps to stay comparable across participants and conditions. 
The second part stated a region goal under the protection constraint of brightening or restoring the subject so its color reads clearly, without changing the background. By stating the region goal in the brief guaranteed that every participant performed the selection-aware, region-scoped work the pre-study found.
The third part was open-ended and creative, asking participants to develop the image toward a final look of their own choosing.

\subsection{Study Procedure}

The study was conducted remotely over Zoom and lasted 60 to 90 minutes. We asked participants to think aloud, collected screen and audio recordings and logged system events (with consent). To prevent order effects we counterbalanced with a 2x2 latin square (Photo x Interface Condition). The study protocol followed four steps:

\begin{enumerate}
    \item \textbf{Introduction:} First participants received a consent form and a short demographics questionnaire. Then a recorded introduction video of the tool was shown that gave every participant the same walkthrough of the editor.
    \item \textbf{Photo-editing tasks:} (2x20 minutes) The participants were asked to edit two different images, with a three part brief to cover different use cases with the tool (\autoref{app:ms-stimuli}). For this, the participants edited one photo with the baseline and the other with the AI layer on, in a counterbalanced order. Each block presented one photograph and the same three-part brief, following the format of a plausible editing session from a global correction to a region-scoped fix to an open editorial finish. The participants were asked to work as they normally would. The brief order was fixed, because the parts built on one another. After each block, the participant completed the per-condition questionnaire. 
    \item \textbf{Comparative Questionnaire:} Participants answered 7 items that compared the two systems (Appendix \ref{app:ms-final}).
    \item \textbf{Semi-structured Interview:} A semi-structured interview was conducted that followed a six-theme guide following the research questions. The themes included: \textit{interface and widget affordances, creativity and exploration, control, trust and authorship, workspace and history, and reflection.}
\end{enumerate}

\subsection{Data collection and analysis}
\label{sec:data-collection}
During the study, we collected screen and audio recordings from the think aloud tasks and the interviews. We also collected questionnaire responses, using NASA-TLX~\cite{hart_development_1988,hart_nasatlx_2006}, System Usability Score~\cite{brooke_sus_1996}, and Creativity Support Index (CSI)~\cite{cherry_quantifying_2014}. We also defined a set of trust and control items grounded in the trust-in-automation constructs identified by Lee and See~\cite{lee_trust_2004} (Appendix \ref{app:ms-trust}), 
and a direct between-condition contrast, which we self-defined (see Appendix \ref{app:ms-final}). We also recorded an event journal where we logged time and distinct operations for every part. 
The exit interview covered questions about the interface, creativity, authorship, workspace and reflection.
We analyzed the data as follows:
\begin{itemize}[leftmargin=*]
    \item \textit{Questionnaire Data}: The paired measures were computed under paired t-test or Wilcoxon signed-rank tests with effect sizes, depending on the normality computed with Shapiro-Wilk on the paired differences. 
    \item \textit{Logged System Data:} The interaction logs were analyzed between the two conditions, comparing time on task, individual parameter-value writes per minute through manual and AI, and mask use. The interaction events included widgets created, accepted, deleted, updated, prompts entered, and compare hold. In the AI-composed conditions descriptive statistics were computed for analysis, analyzing widgets creation, coexistent widgets, AI acceptance rate and refinement behavior. 
    \item \textit{Interview Data:} The recordings were transcribed and translated, as the interviews were conducted in German. For this, we used Whisper from OpenAI locally on a secure computer~\cite{radford2023robust}. Our research team is fluent in English and German, which allowed us to ensure an accurate translation. We listened to the recordings and checked the correctness of the translation. We analyzed the transcripts using a deductive thematic analysis, as in the formative study (Section \ref{sec:formative-study}), applying the methodology from Braun and Clarke~\cite{braun_using_2006, braun_reflecting_2019,braun_one_2021}. 
\end{itemize}

\section{Results}
In this section, we present the results of our mixed-methods study. We start by reporting the descriptive statistics from the questionnaire. Then we present the results of the analysis of the logged system data. Finally, we present the results of the thematic analysis, which allows us to get a detailed understanding of the participants experience with the tool. 

\subsection{Questionnaire analysis}


\subsubsection{Usability and workload were comparable, but participants perceived less mental effort with AI composition}

The SUS scores (see \autoref{fig:sus-boxplot}) and the NASA-TLX results did not reach significance between the conditions, and no individual NASA-TLX item differed significantly between conditions (\autoref{tab:tlx}). The final comparison survey however showed a clearer subjective preference regarding mental effort: six participants felt that the AI-composed version required less mental effort, three reported no difference, and three favored the Baseline (\autoref{fig:results-fc}). Thus, while the workload measures did not provide evidence of a significant reduction, participants more often perceived the AI-composed workflow as requiring less mental effort when directly comparing the two versions.

\subsubsection{Exploration and enjoyment were higher with AI composition}

The overall CSI score was not statistically significant (\autoref{tab:csi}). At the factor level, \textbf{Exploration} ($p=.024$) and \textbf{Enjoyment} ($p=.049$) were significantly higher with AI composition, while Expressiveness, Immersion, and Results Worth Effort showed no significant differences between conditions. The final comparison was consistent with the higher Exploration scores. Eight of twelve participants felt that the AI-composed version helped them explore ideas or be more creative, while four reported no difference and none favored the Baseline (\autoref{fig:results-fc}). These results indicate that the clearest creativity-related benefit of AI composition concerned exploring possibilities and enjoying the editing process. 

\subsubsection{Participants preferred AI composition despite some loss of control and authorship}

\begin{figure*}[htbp]
\centering

\definecolor{L1}{HTML}{B2182B}
\definecolor{L2}{HTML}{EF8A62}
\definecolor{L3}{HTML}{FDDBC7}
\definecolor{L4}{HTML}{E0E0E0}
\definecolor{L5}{HTML}{D1E5F0}
\definecolor{L6}{HTML}{67A9CF}
\definecolor{L7}{HTML}{2166AC}

\begin{tikzpicture}

\def\barheight{0.34}
\def\labelx{-2.70}

\def\labelfont{\small}
\def\countfont{\scriptsize}
\def\directionfont{\scriptsize}

\def\yPref{0.00}
\def\yControl{-0.65}
\def\yTrust{-1.30}
\def\yExplore{-1.95}
\def\yEffort{-2.60}
\def\ySatisfied{-3.25}
\def\yAuthorship{-3.90}

\draw[black!45, dashed]
    (0,-4.15) -- (0,0.35);

\fill[L4]
    (-0.17,\yPref-\barheight/2)
    rectangle
    (0.17,\yPref+\barheight/2);

\node[font=\countfont, text=black!75]
    at (0,\yPref) {1};

\fill[L6]
    (0.17,\yPref-\barheight/2)
    rectangle
    (3.91,\yPref+\barheight/2);

\node[font=\countfont, text=black!75]
    at (2.04,\yPref) {11};

\node[anchor=east, font=\labelfont]
    at (\labelx,\yPref) {Preferred version};

\fill[L1]
    (-2.55,\yControl-\barheight/2)
    rectangle
    (-1.87,\yControl+\barheight/2);

\node[font=\countfont, text=black!75]
    at (-2.21,\yControl) {2};

\fill[L2]
    (-1.87,\yControl-\barheight/2)
    rectangle
    (-0.51,\yControl+\barheight/2);

\node[font=\countfont, text=black!75]
    at (-1.19,\yControl) {4};

\fill[L4]
    (-0.51,\yControl-\barheight/2)
    rectangle
    (0.51,\yControl+\barheight/2);

\node[font=\countfont, text=black!75]
    at (0,\yControl) {3};

\fill[L6]
    (0.51,\yControl-\barheight/2)
    rectangle
    (0.85,\yControl+\barheight/2);

\node[font=\countfont, text=black!75]
    at (0.68,\yControl) {1};

\fill[L7]
    (0.85,\yControl-\barheight/2)
    rectangle
    (1.53,\yControl+\barheight/2);

\node[font=\countfont, text=black!75]
    at (1.19,\yControl) {2};

\node[anchor=east, font=\labelfont]
    at (\labelx,\yControl) {Felt in control};

\fill[L1]
    (-2.21,\yTrust-\barheight/2)
    rectangle
    (-1.87,\yTrust+\barheight/2);

\node[font=\countfont, text=black!75]
    at (-2.04,\yTrust) {1};

\fill[L2]
    (-1.87,\yTrust-\barheight/2)
    rectangle
    (-0.85,\yTrust+\barheight/2);

\node[font=\countfont, text=black!75]
    at (-1.36,\yTrust) {3};

\fill[L4]
    (-0.85,\yTrust-\barheight/2)
    rectangle
    (0.85,\yTrust+\barheight/2);

\node[font=\countfont, text=black!75]
    at (0,\yTrust) {5};

\fill[L6]
    (0.85,\yTrust-\barheight/2)
    rectangle
    (1.19,\yTrust+\barheight/2);

\node[font=\countfont, text=black!75]
    at (1.02,\yTrust) {1};

\fill[L7]
    (1.19,\yTrust-\barheight/2)
    rectangle
    (1.87,\yTrust+\barheight/2);

\node[font=\countfont, text=black!75]
    at (1.53,\yTrust) {2};

\node[anchor=east, font=\labelfont]
    at (\labelx,\yTrust) {Trusted the tool};

\fill[L4]
    (-0.68,\yExplore-\barheight/2)
    rectangle
    (0.68,\yExplore+\barheight/2);

\node[font=\countfont, text=black!75]
    at (0,\yExplore) {4};

\fill[L6]
    (0.68,\yExplore-\barheight/2)
    rectangle
    (2.72,\yExplore+\barheight/2);

\node[font=\countfont, text=black!75]
    at (1.70,\yExplore) {6};

\fill[L7]
    (2.72,\yExplore-\barheight/2)
    rectangle
    (3.40,\yExplore+\barheight/2);

\node[font=\countfont, text=black!75]
    at (3.06,\yExplore) {2};

\node[anchor=east, font=\labelfont]
    at (\labelx,\yExplore) {Explore ideas / creativity};

\fill[L2]
    (-1.53,\yEffort-\barheight/2)
    rectangle
    (-0.51,\yEffort+\barheight/2);

\node[font=\countfont, text=black!75]
    at (-1.02,\yEffort) {3};

\fill[L4]
    (-0.51,\yEffort-\barheight/2)
    rectangle
    (0.51,\yEffort+\barheight/2);

\node[font=\countfont, text=black!75]
    at (0,\yEffort) {3};

\fill[L6]
    (0.51,\yEffort-\barheight/2)
    rectangle
    (1.19,\yEffort+\barheight/2);

\node[font=\countfont, text=black!75]
    at (0.85,\yEffort) {2};

\fill[L7]
    (1.19,\yEffort-\barheight/2)
    rectangle
    (2.55,\yEffort+\barheight/2);

\node[font=\countfont, text=black!75]
    at (1.87,\yEffort) {4};

\node[anchor=east, font=\labelfont]
    at (\labelx,\yEffort) {Less mental effort};

\fill[L1]
    (-1.36,\ySatisfied-\barheight/2)
    rectangle
    (-1.02,\ySatisfied+\barheight/2);

\node[font=\countfont, text=black!75]
    at (-1.19,\ySatisfied) {1};

\fill[L2]
    (-1.02,\ySatisfied-\barheight/2)
    rectangle
    (-0.68,\ySatisfied+\barheight/2);

\node[font=\countfont, text=black!75]
    at (-0.85,\ySatisfied) {1};

\fill[L4]
    (-0.68,\ySatisfied-\barheight/2)
    rectangle
    (0.68,\ySatisfied+\barheight/2);

\node[font=\countfont, text=black!75]
    at (0,\ySatisfied) {4};

\fill[L6]
    (0.68,\ySatisfied-\barheight/2)
    rectangle
    (2.04,\ySatisfied+\barheight/2);

\node[font=\countfont, text=black!75]
    at (1.36,\ySatisfied) {4};

\fill[L7]
    (2.04,\ySatisfied-\barheight/2)
    rectangle
    (2.72,\ySatisfied+\barheight/2);

\node[font=\countfont, text=black!75]
    at (2.38,\ySatisfied) {2};

\node[anchor=east, font=\labelfont]
    at (\labelx,\ySatisfied) {Satisfied with image};

\fill[L1]
    (-2.55,\yAuthorship-\barheight/2)
    rectangle
    (-1.87,\yAuthorship+\barheight/2);

\node[font=\countfont, text=black!75]
    at (-2.21,\yAuthorship) {2};

\fill[L2]
    (-1.87,\yAuthorship-\barheight/2)
    rectangle
    (-1.19,\yAuthorship+\barheight/2);

\node[font=\countfont, text=black!75]
    at (-1.53,\yAuthorship) {2};

\fill[L4]
    (-1.19,\yAuthorship-\barheight/2)
    rectangle
    (1.19,\yAuthorship+\barheight/2);

\node[font=\countfont, text=black!75]
    at (0,\yAuthorship) {7};

\fill[L6]
    (1.19,\yAuthorship-\barheight/2)
    rectangle
    (1.53,\yAuthorship+\barheight/2);

\node[font=\countfont, text=black!75]
    at (1.36,\yAuthorship) {1};

\node[anchor=east, font=\labelfont]
    at (\labelx,\yAuthorship) {Sense of authorship};

\def\ydirection{-4.40}

\node[anchor=east, font=\directionfont]
    at (-0.15,\ydirection)
    {$\leftarrow$ favors Baseline};

\node[anchor=west, font=\directionfont]
    at (0.15,\ydirection)
    {favors AI-composed $\rightarrow$};

\def\ylegend{-4.92}
\def\boxw{0.30}
\def\boxh{0.22}


\fill[L1]
    (-2.16-\boxw/2,\ylegend-\boxh/2)
    rectangle
    (-2.16+\boxw/2,\ylegend+\boxh/2);

\fill[L2]
    (-1.08-\boxw/2,\ylegend-\boxh/2)
    rectangle
    (-1.08+\boxw/2,\ylegend+\boxh/2);

\fill[L4]
    (0-\boxw/2,\ylegend-\boxh/2)
    rectangle
    (0+\boxw/2,\ylegend+\boxh/2);

\fill[L6]
    (1.08-\boxw/2,\ylegend-\boxh/2)
    rectangle
    (1.08+\boxw/2,\ylegend+\boxh/2);

\fill[L7]
    (2.16-\boxw/2,\ylegend-\boxh/2)
    rectangle
    (2.16+\boxw/2,\ylegend+\boxh/2);

\end{tikzpicture}

\caption[Final comparison, $N=12$]
{Final comparison, $N=12$. Each bar shows how many participants favored each version: the top bar shows overall preference, while the six bars below show the comparative items. Bars diverge around the neutral category, such that responses to the right favor the AI-composed version and responses to the left favor the baseline. Segment labels indicate participant counts.}
\Description{Diverging stacked bar chart, N=12, showing preference ratings for seven measures on a five-point scale, with segments running from dark red (strongly favors Baseline) through gray (neutral) to dark blue (strongly favors AI-composed); full counts are listed in Appendix \ref{app:results} in Table \ref{fig:results-fc}. Every measure leans toward the AI-composed side of neutral, most sharply for Preferred version, where 11 of 12 participants favored AI-composed against 1 neutral response and zero favoring Baseline. Explore ideas / creativity also has no Baseline-favoring responses at all. Sense of authorship is the most neutral measure, with 7 of 12 responses in the neutral category and none strongly favoring either version.}
\label{fig:results-fc}
\end{figure*}
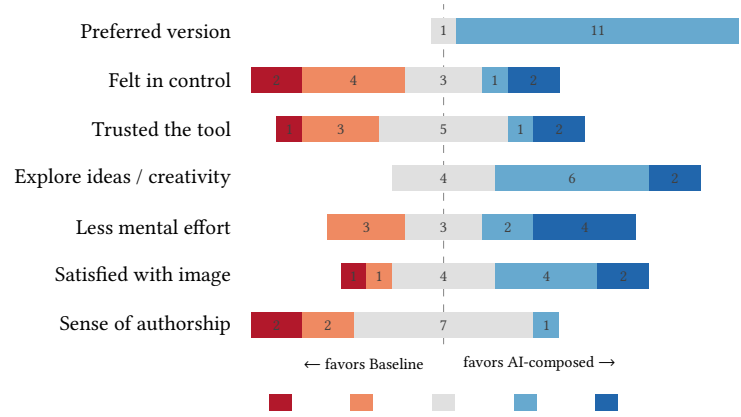

The final comparison (\autoref{fig:results-fc}) reflected this tension. Eleven of twelve participants preferred the AI-composed version overall, while one rated the two versions equally and none preferred the Baseline. Yet this preference did not mean that AI composition increased participants' sense of control or authorship. Six of twelve participants felt more in control with the Baseline, compared with three who favored AI composition and three who rated the conditions equally. This pattern was also reflected in perceived control (\autoref{tab:trust}), which was significantly lower in the AI-composed condition ($p = 0.046$). Four participants felt that the final image was more their own work with the Baseline, compared with one who favored AI composition; seven perceived no difference in authorship. At the same time, participants favored AI composition on other aspects of the experience, including satisfaction with the final image. Future-use responses followed the same pattern. Nine of twelve participants wanted to use AI composition for corrective edits, and nine wanted to use it for open-ended creative edits. These findings show that participants' overall preference for AI composition did not depend on feeling more in control or experiencing stronger authorship. Instead, they appeared willing to accept some reduction in these qualities when AI composition provided sufficient benefits elsewhere.

\subsubsection{Exploratory differences by expertise level}

An exploratory analysis by expertise ($N=6$ intermediate, $N=6$ professional) found few clear differences between the two groups. None of the combined measures, including SUS, NASA-TLX, and CSI, showed a significant difference in how the two groups responded to AI composition. 


At the factor level, mental demand was the only measure that differed nominally significantly by expertise ($p=.027$): it decreased with AI composition for intermediate participants, while changing little for professionals (\autoref{tab:rtlx-expertise}, Appendix). No CSI factor showed a significant difference between expertise groups, although overall CSI was descriptively higher with AI assistance in both groups (intermediate: \(M_{\text{Baseline}}=74.17, SD=18.33\) vs. \(M_{\text{AI-composed}}=77.83, SD=13.13\); professional: \(M_{\text{Baseline}}=73.30, SD=16.24\) vs. \(M_{\text{AI-composed}}=82.83, SD=8.35\)), with a larger descriptive increase among professionals (\autoref{tab:csi-expertise}, Appendix). Given the small groups ($N=6$ each) and exploratory nature of these comparisons, these patterns should be interpreted cautiously. Overall, the results provide limited evidence for expertise-related differences, but suggest that experience may influence which aspects of AI composition users benefit from.

\begin{table*}[htbp]
\centering
\caption{Creativity Support Index (CSI): AI-composed vs.\ Baseline ($N=12$). Bold marks $p<.05$. As our study did not involve collaborative tasks among humans, we omitted the collaboration factor, following prior work~\cite{carollCreativityFactorEvaluation2009, codetoon, amin_promptcanvas_2025}.}
\label{tab:csi}
\resizebox{\textwidth}{!}{%
\begin{tabular}{lccccccccccc}
\toprule
& \multicolumn{2}{c}{AI-composed} & \multicolumn{2}{c}{Baseline} & \multicolumn{1}{c}{AI-composed} & \multicolumn{1}{c}{Baseline} & & & \multicolumn{3}{c}{Statistics} \\
\cmidrule(lr){2-3} \cmidrule(lr){4-5} \cmidrule(lr){6-6} \cmidrule(lr){7-7} \cmidrule(lr){10-12}
Measure & $M$ & $SD$ & $M$ & $SD$ & $Mdn (IQR)$ & $Mdn (IQR)$ & Shapiro $p$ & Test & $p$ & Effect size & 95\% CI \\
\midrule
Expressiveness & 16.08 & 3.20 & 13.92 & 4.87 & 16.50 (16.00--17.25) & 14.00 (11.50--18.00) & 0.045 & Wilcoxon & 0.178 & $r = 0.45$ & $[-1.00, 4.00]$ \\
Exploration & 16.58 & 3.09 & 13.75 & 4.73 & 18.00 (15.00--19.00) & 13.50 (11.25--17.50) & 0.487 & Paired $t$ & \textbf{0.024} & $d_z = 0.75$ & $[0.44, 5.22]$ \\
Immersion & 13.25 & 4.03 & 12.50 & 4.50 & 13.50 (10.75--15.75) & 14.00 (8.00--15.25) & 0.269 & Paired $t$ & 0.646 & $d_z = 0.14$ & $[-2.74, 4.24]$ \\
Enjoyment & 17.25 & 2.49 & 15.58 & 3.34 & 18.00 (16.00--18.25) & 15.50 (13.00--19.00) & 0.046 & Wilcoxon & \textbf{0.049} & $r = 0.78$ & $[0.00, 3.50]$ \\
Results worth effort & 16.17 & 2.72 & 15.67 & 2.53 & 16.00 (14.75--17.50) & 15.50 (14.00--17.25) & 0.215 & Paired $t$ & 0.609 & $d_z = 0.15$ & $[-1.59, 2.59]$ \\
\midrule
Overall CSI & 80.33 & 10.81 & 73.73 & 16.52 & 82.50 (74.50--86.00) & 72.25 (65.62--85.12) & 0.644 & Paired $t$ & 0.136 & $d_z = 0.46$ & $[-2.44, 15.64]$ \\
\bottomrule
\end{tabular}%
}
\begin{minipage}{\textwidth}
\vspace{2pt}
\footnotesize \textit{Note.} $N=12$. Test per row selected via Shapiro--Wilk on the paired differences
(AI-composed $-$ Baseline) at $\alpha=.05$: paired $t$-test (Cohen's $d_z$) where normal, Wilcoxon signed-rank
test (matched-pairs rank-biserial $r$) otherwise. CIs are on the mean difference for paired-$t$ rows and
the Hodges--Lehmann estimate of the median difference (bootstrap, 10{,}000 resamples) for Wilcoxon rows.
\end{minipage}
\end{table*}

\subsection{System data analysis}
\label{subsection:systemloganalysis}

Interaction logs allowed us to analyze how participants used the AI-composed widgets, complementing their self-reports (\autoref{tab:paired_comparisons}). Time on task did not differ significantly between conditions (AI-composed: $M=16.57$ min; Baseline: $M=16.82$ min; $p=.791$). However, interaction density differed substantially. Total interaction events increased from $13.34$ to $39.43$ events per minute ($p<.001$), while participant-authored parameter edits increased from $10.98$ to $26.15$ edits per minute ($p=.001$). This suggests that AI composition increased hands-on editing activity rather than replacing it.

Focusing on behavior within the AI-composed condition (\autoref{tab:results_ai_on}), participants created an average of 10.58 widgets per session ($SD=5.05$). These were distributed between manually invoked widgets ($M=4.58$), prompted AI proposals ($M=4.50$), and autonomous AI proposals ($M=1.50$), indicating that explicit user invocation was substantially more common than autonomous suggestion. Participants also kept several widgets active simultaneously, with an average session maximum of 6.58 coexistent widgets. AI proposals were accepted without modification at a mean rate of .35 ($Mdn=.23$, $SD=.38$). Participants accepted an average of 2.00 proposals and dismissed 2.58. 
Refinement was also frequent, with an average of 18.33 refinement episodes per session and 7.58 refinement actions per episode. These behavioral patterns suggest that participants primarily treated AI-composed controls as starting points for further manipulation rather than as finished outputs.

\begin{table*}[t]
    \centering
    \caption{Paired comparisons of session behavior between the
    \textit{AI-composed} and \textit{Baseline} conditions ($n=12$).}
    \label{tab:paired_comparisons}

    \renewcommand{\arraystretch}{0.95}
    \setlength{\tabcolsep}{5pt}

    \begin{threeparttable}
        \begin{tabular}{@{}lrrrrr@{}}
            \toprule
            \textbf{Measure} &
            \multicolumn{1}{c}{\textbf{AI-composed}} &
            \multicolumn{1}{c}{\textbf{Baseline}} &
            \multicolumn{1}{c}{$V$} &
            \multicolumn{1}{c}{$p$} &
            \multicolumn{1}{c}{$r_{rb}$} \\
            &
            \multicolumn{1}{c}{\textbf{\textit{M (SD)}}} &
            \multicolumn{1}{c}{\textbf{\textit{M (SD)}}} &
            & & \\
            \midrule

            Duration (min)
            & 16.57 (8.06)
            & 16.82 (6.34)
            & 43.0
            & .791
            & .10 \\

            Total events / min
            & 39.43 (2.25)
            & 13.34 (5.43)
            & 78.0
            & $<$.001
            & 1.00 \\

            \addlinespace[2pt]
            \multicolumn{6}{@{}l}{\textbf{Edits}} \\

            \hspace{1em}Manual edits / min
            & 26.15 (18.42)
            & 10.98 (3.98)
            & 77.0
            & .001
            & .97 \\

            \hspace{1em}AI edits / min
            & 8.77 (4.04)
            & 0.49 (1.28)
            & 78.0
            & $<$.001
            & 1.00 \\

            \addlinespace[2pt]
            \multicolumn{6}{@{}l}{\textbf{Masks}} \\

            \hspace{1em}Masks created
            & 2.42 (1.51)
            & 1.67 (0.78)
            & 32.5
            & .293
            & .44 \\

            \hspace{1em}Masks deleted
            & 0.75 (0.87)
            & 0.50 (1.00)
            & 14.0
            & .625
            & .33 \\

            \bottomrule
        \end{tabular}

        \begin{tablenotes}[flushleft]
            \footnotesize
            \item[] \textit{Note.}
            $V$ = Wilcoxon signed-rank statistic (sum of positive ranks);
            $r_{rb}$ = matched-pairs rank-biserial correlation.
            Events include widgets created, accepted, deleted, or updated,
            prompts entered, and compare-hold interactions.
            Edits count parameter changes made either by the participant or
            by the AI, such as accepting a proposal or refinement.
        \end{tablenotes}

    \end{threeparttable}

    \Description{Paired comparisons of session behavior between the
    AI-composed and Baseline conditions. AI-composed sessions had more
    total events per minute, more manual edits per minute, and more AI
    edits per minute, while session duration and mask creation and
    deletion did not differ significantly between conditions.}
\end{table*}

\begin{table}[t]
    \centering
    \caption{Descriptive statistics for widget creation, AI proposal outcomes, 
    and refinement behavior during \textit{AI-composed} sessions ($n=12$).}
    \label{tab:results_ai_on}

    \renewcommand{\arraystretch}{0.95}
    \setlength{\tabcolsep}{5pt}

    \begin{threeparttable}

        \begin{tabular}{@{}lrrr@{}}
            \toprule
            \textbf{Measure} &
            \textit{\textbf{M}} &
            \textit{\textbf{SD}} &
            \textit{\textbf{Mdn (IQR)}} \\
            \midrule

            Widgets created
            & 10.58 & 5.05 & 10.5 (7.5--13.0) \\

            \hspace{1em}Autonomous
            & 1.50 & 2.11 & 0.5 (0.0--2.25) \\

            \hspace{1em}Prompted
            & 4.50 & 3.09 & 4.5 (2.5--6.5) \\

            \hspace{1em}Manual
            & 4.58 & 4.48 & 3.5 (0.0--9.25) \\

            \addlinespace[2pt]

            Coexistent widgets (max)
            & 6.58 & 3.26 & 6.5 (4.75--8.0) \\

            \addlinespace[2pt]

            \multicolumn{4}{@{}l}{\textbf{AI proposals}} \\

            \hspace{1em}Acceptance rate
            & 0.35 & 0.38 & 0.23 (0.0--0.59) \\

            \hspace{1em}Accepted
            & 2.00 & 2.63 & 1.0 (0.0--2.75) \\

            \hspace{1em}Dismissed
            & 2.58 & 3.09 & 1.5 (0.0--4.25) \\

            \addlinespace[2pt]

            \multicolumn{4}{@{}l}{\textbf{Refinement behavior}} \\

            \hspace{1em}Refine episodes
            & 18.33 & 13.72 & 14.5 (7.75--25.0) \\

            \hspace{1em}Refines / episode
            & 7.58 & 2.30 & 8.30 (5.67--8.99) \\

            \bottomrule
        \end{tabular}

        \begin{tablenotes}[flushleft]
            \footnotesize
            \item[] \textit{Note.}
            Acceptance rate is the proportion of AI proposals accepted without
            modification. A refine episode represents a burst of refinement
            activity; refines per episode indicates the number of refinement
            actions performed within each episode.
        \end{tablenotes}

    \end{threeparttable}

    \Description{Descriptive statistics for widget creation, AI proposal
    outcomes, and refinement behavior during AI-composed sessions.
    Widgets were created through autonomous AI proposals, prompted AI
    proposals, and manual creation. Participants accepted approximately
    one third of AI proposals without modification. Refinement involved
    multiple episodes with several refinement actions per episode.}
\end{table}

\subsection{Interview analysis}
The preliminary themes we got from our thematic analysis included: \textit{interface and widget affordances, creativity and exploration, control, trust and authorship, workspace and history, and reflection.} Through the deductive analysis, we unified the themes \textit{creativity and exploration, Efficiency and Navigating Editing Complexity, Control, and Authorship and AI Involvement}.

\subsubsection{Creativity and Exploration}
Our findings indicate that the AI-composed parametric controls inspired experimentation and unexpected directions. However this guidance also was perceived as steering and constraining exploration, being phrased as dilemma of expanding but also limiting creativity.

\paragraph{AI suggestions encouraged experimentation and unexpected directions}

Participants described AI-composed controls as providing starting points they could experiment with rather than fixed outcomes they had to accept, tying in with subsection \ref{subsection:systemloganalysis}. P10 captured this exploratory attitude directly: \textit{``I'm allowed to mess around, right? [...] I can just go ahead and click on things and see what happens''} (P10). Unexpected results could themselves become new creative directions. When an AI-generated edit replaced the requested grass background with a Swedish lake, P10 responded, \textit{``Oh, oh, I didn't expect that. But now we're just in Sweden. I think that's great''} (P10). P2 similarly became more playful as he grew familiar with the system, shifting from immediately solving the task toward exploring what the AI might suggest.

This meant that accuracy was not always the only criterion by which participants judged AI suggestions. P1 estimated that suggestions were only \textit{``about 50 percent or maybe 60 percent''} accurate, but considered this acceptable during exploration because \textit{``that's what you want -- to be able to try out different things. And maybe you also want to see what isn't right''}. Participants also valued that this experimentation remained open to further manipulation. P11 described the interface as providing \textit{``a lot of room to work with''} and allowing users to \textit{``really let your creativity run wild''}. Thus, suggestions could be useful not only when they correctly realized an existing intention, but also when they provided alternatives that participants could explore, reject, or develop further.

\paragraph{AI guidance could steer and constrain exploration}

The same suggestions that expanded the space of possibilities could also shape which possibilities participants considered. P1 described this explicitly: \textit{``you're more likely to accept those suggestions if they fit the project -- even if I wouldn't have come up with them myself''} (P1). P4 also mentioned that the AI suggestions could implicitly shape the creative direction: \textit{``because you’re just working with what you have in front of you''}. P5, one of the intermediate participants, saw this as an opportunity as the AI suggestions could help participants continue experimenting when they became stuck.
P9 experienced a similar shift of the AI guidance. With AI composition, he became more absorbed in exploring \textit{``what options the AI was actually suggesting''} and felt \textit{``a little more creative''}, but also acknowledged, \textit{``I got a little hooked on what the AI was suggesting to me. And then I used my brain a little less''} (P9). 
P1 characterized the effect as a dilemma: \textit{``[the creativity was] not really limited, but it’s just being guided in a certain direction [...] And of course, it’s also expanded it. So it’s just a bit of a dilemma [...] It goes both ways--a bit of a positive and a negative.''}.

This can be a problem if participants want to prompt unexpected or non-normative things. P3 for example wanted to change the color of the boat to green, but struggled with it. She talked about that \textit{``the AI understood what I wanted and which object. But there were some issues with the execution. It understood `brighter and more natural' pretty well—brighter, warmer, and things like that. But when I wanted the boat to be green, the boat didn’t turn green''}. 

This is important, as the AI suggestions guide the edits and also prime the user, P11 put it: \textit{``Even if you just catch a brief glimpse of an image, like the one the AI had in mind, you’re already being primed very quickly''}. This is something they did not want, because they wanted to bring their own visual language and thinking into the image. Therefore, they continued to value the option to step away from AI guidance altogether. This also allowed them to work \textit{``old-school''}, like working in \textit{``the `Photoshop 2006' way, so to speak -- build your own stuff and then you have your own result, which is completely in your hands''} (P11). So, AI composition 
changed which possibilities participants encountered and where those possibilities came from, while making self-directed editing an important alternative.

\subsubsection{Efficiency and Navigating Editing Complexity}
Our results suggest that the AI-composed widgets helped with navigating the photo editor, making the use more efficient for both intermediates and experts. The widgets also helped with making editing operations easier to understand and to refine.

\paragraph{AI-composed controls reduced search and trial-and-error}

Participants described AI-composed widgets as reducing the effort required to find suitable editing operations. This was seen as particularly valuable for less-experienced users, who might otherwise need to know which parameters to search for. P2 explained that being able to describe an intended change and receive relevant controls could lower the barrier to entry because users \textit{``just don't have to search through 400 sliders''} (P2). The same mechanism was perceived as reducing trial-and-error for more experienced editing workflows. P10 described the generated controls as \textit{``faster than figuring it out on your own [...] saving you from trial and error''} (P10). In some cases, the system also surfaced adjustments participants would not have readily discovered themselves. Reflecting on an AI-suggested color adjustment, P1 explained, \textit{``I wouldn't have figured that out on my own, that adding orange to water makes it look naturally bluer [...] that would've taken forever, and this way was right there, so it went much faster''} (P1). So, AI composition supported efficiency by automating edits as well as by bringing potentially relevant editing operations into reach without requiring participants to search for or discover them manually.

\paragraph{Widgets made editing operations easier to understand and refine}

Once generated, participants valued that AI-composed edits remained visible and adjustable rather than being applied as fixed results. P1 appreciated being able to inspect individual suggestions and decide whether to keep them: \textit{``I don't have to accept them and [...] can turn it on and off again, so that I know exactly what the adjustments are''} (P1). Sensible defaults also provided a useful starting point without preventing further refinement. P3 explained, \textit{``it's great that there are default settings already in place [...] But of course, sometimes I want to tweak it, and I was able to do that too''} (P3).

\subsubsection{Control, Authorship, and AI Involvement}
\label{sec:control}
One aim of this research was to understand the impact of the AI-composed, manipulable controls on the perceived control and authorship of the participants. We found during the interviews, that participants located the authorship in how they directed and refined the AI-composed controls themselves. The possibility to adjust and reverse also preserved the perceived control, however this had its limits. 

\paragraph{Participants located authorship in how they directed and refined AI}

Participants generally did not describe AI-assisted editing as either entirely their own work or entirely AI-generated. Instead, authorship depended on how the AI contributed and on participants' ability to decide what remained in the final image. P10 described AI as \textit{``a tool I use''}, emphasizing that \textit{``the AI does not choose the photo or the type of editing''} (P10). Even when the AI introduced unexpected content, such as replacing the requested grass with a Swedish lake, they still located the final decision with herself: \textit{``at the end of the day, I'm the one who decides whether I want to keep it that way or not''}. So, they characterized the resulting image as \textit{``a mix -- it's a co-op''} (P10).

However, participants also distinguished between different kinds of AI involvement. P7 did not consider AI adjustment of existing parameters such as exposure or contrast to constitute meaningful co-creation, drawing a distinction between parameter-level assistance and changes to the actual image content. An important part of this distinction was that users could manually refine AI suggestions. Rather than repeatedly prompting the AI to correct an edit, they valued that \textit{``here I can just adjust the sliders or the numbers myself, to fine-tune it exactly to my preferences''} (P7). 

\paragraph{Adjustability and reversibility preserved control, but had limits}

Participants felt more comfortable with AI involvement when its suggestions remained adjustable, reversible, and optional. P9, for example, did not feel pushed toward the AI's suggestions because \textit{``I could just remove them again''}. Similarly, P10 described AI assistance as \textit{``a task I assign''} while emphasizing that \textit{``I still retain control over that task''} (P10). In this sense, delegating part of the editing process to AI did not necessarily mean giving up control, as long as participants could subsequently inspect, adjust, or reject its contribution. However, this control depended on the available range of adjustment. P2 found the suggested defaults reasonable, but in some cases the controls did not allow her to push an effect far enough: \textit{``when I turned the sliders all the way up, the effect didn't go far enough''} (P2). She also recognized that the relevance of generated controls depended partly on how clearly she communicated her intentions, noting that \textit{``it can only make suggestions as good as how precisely I ask''} (P2). Thus, manual refinement provided a way to recover from imperfect suggestions. 

Beyond adjusting individual suggestions, some participants wanted control over whether AI intervened at all. P11 emphasized that it was \textit{``important that you can turn it off completely''} (P11), reflecting a concern that continuously presented AI suggestions could subtly steer the editing process. Participants valued control at multiple levels: specifying their intentions, refining or rejecting individual outputs, and deciding when AI assistance entered the workflow in the first place.

\section{Discussion and Future Work}
\label{sec:discussion}

In this work, we investigated how AI-assisted image-editing interfaces can translate high-level editing intentions into directly manipulable parametric controls. Specifically, we examined (RQ1) how such intentions can be translated into context-specific parametric controls that remain open to refinement, (RQ2) how this translation affects practitioners' editing experience and creative exploration, and (RQ3) how AI involvement in selecting and initializing controls affects practitioners' felt control and sense of authorship. Across these questions, our findings suggest that semantic interface generation changes the role of AI in image editing. Rather than only producing an edited result, AI can participate in constructing the space of actions through which users continue to work. This can make relevant operations easier to access and provide starting points for exploration, but it also gives the system influence over which possibilities become visible and salient. We discuss these implications below.

\subsection{Translating Intentions into Refinable Parametric Controls}
\label{sec:discussion-translation}

Addressing \textbf{RQ1}, our findings suggest that high-level editing intentions are most useful when translated into parametric controls as starting points for refinement rather than completed edits. Atelier interprets an intention in the context of the image and proposes relevant operations, spatial targets, and initial values that remain inspectable and adjustable. Participant behavior supports this approach. In the AI-composed condition, interaction density nearly tripled and participant-authored parameter adjustments more than doubled, while task completion time did not differ significantly. Participants continued to refine, supplement, and reject AI proposals. So, the system helped identify relevant operations without replacing direct manipulation. This extends prior work on graphical interaction with generative systems \cite{davis_fashioning_2024}. Rather than responding to an unsatisfactory result by reformulating a prompt \cite{swift_value-oriented_2024}, users can directly adjust the system's interpretation. Importantly, the proposed value need not be correct for the translation to be useful: a temperature control may appropriately represent ``make this warmer'' even if its initial setting is too strong.

Similar to DesignTrace~\cite{peng_designtrace_2026}, Atelier keeps intermediate representations available for continued interaction. However, rather than maintaining a semantic substrate of persistent design attributes, Atelier introduces a \emph{parametric substrate}: semantic intent is translated into familiar editing operations and values, such as exposure, white balance, curves, or color adjustments. Because AI-composed and manual controls share the same operation registry and editing state, users can move between AI assistance and conventional editing without switching modes.

The translation can also specify \emph{where} an operation applies through persistent spatial targets. While this may help users track which controls affect which regions, our NASA-TLX results do not demonstrate reduced cognitive demand, and the Baseline offered the same segmentation capabilities. We therefore leave this as a direction for further study \cite{metacognition}.

So, Atelier suggests one approach to semantic-to-parametric translation. Identify a small set of plausible domain-native operations, initialize them as useful starting points, and keep both their values and their composition open to further manipulation. Instead of hiding technical editing behind a semantic command, semantic reasoning can help users reach the technical operations through which their intention can be realized.

\subsection{AI Suggestions Both Expand and Shape Exploration}
\label{sec:discussion-exploration}

Addressing \textbf{RQ2}, AI composition affected exploration. Exploration and Enjoyment were significantly higher with AI-composed controls. No participant preferred the Baseline for supporting exploration. AI suggestions supported exploration by turning possible interpretations into controls participants could immediately try. Even unexpected suggestions become useful starting points, exposing directions participants might not otherwise have considered. Interaction could help users develop their intentions instead of requiring those intentions to be fully formed beforehand. However, AI suggestions also shaped exploration. Participants sometimes became absorbed in proposed options or relied less on their own reasoning. Choosing which controls to surface therefore directs attention toward some possibilities while making others less visible. Higher exploration should consequently not be interpreted as necessarily more self-directed or comprehensive exploration. Simply exposing more options is unlikely to solve this tension, since additional controls can increase interaction difficulty \cite{dang_ganslider_2022}. Instead, semantic interfaces could help users move between AI-guided and self-directed exploration, for example by offering contrasting interpretations or preserving easy access to the full conventional toolset.

Rejection can also inform this process. Rather than treating a rejected suggestion only as a failed prediction, systems could use it to infer users' authorial boundaries \cite{fu_vistoria_2026} and offer meaningfully different alternatives. Because high-level intentions often admit multiple plausible realizations, semantic interfaces should support alternative interpretations rather than presenting a single translation.

\subsection{Actionability Does Not Guarantee Felt Control}
\label{sec:discussion-control}

Addressing \textbf{RQ3}, our findings reveal a tension between preference and perceived control. Eleven of twelve participants preferred the AI-composed condition, yet participants reported significantly lower felt control with AI composition. At the same time, most did not report a weaker sense of authorship, and participants valued being able to adjust, reject, or remove AI proposals. This aligns with prior work showing that selection and manipulation can preserve agency in human--AI workflows \cite{fu_vistoria_2026,masson_textoshop_2025}. One explanation is that users retained control over \emph{how} AI proposals were adjusted, but had less control over \emph{how those proposals were initially formed}. Atelier gives users substantial execution control over AI proposals, but the AI still interprets the intention, selects relevant operations, and initializes their values. Users can refine what the system proposes while having less influence over how the editing space is initially structured.

Direct manipulation can thus make AI interpretations actionable without making them fully legible. Future semantic interfaces could expose more of the translation process by allowing users to replace operations, redirect targets, request alternative interpretations, or inspect why particular controls were proposed. Following mixed-initiative principles \cite{horvitz_principles_1999}, evaluations of AI-assisted creative tools should therefore distinguish between control over execution, control over framing, and authorship.

\subsection{Limitations and Future Work}
\label{sec:discussion-limitations}

Our findings should be interpreted in light of several limitations. First, the study involved twelve practitioners completing two time-limited tasks in a single session. The small sample, limited task set, and use of think-aloud constrain generalizability. Longitudinal studies with participants' own images are needed to examine whether the observed exploration and refinement patterns persist over time.

Second, the Baseline retained segmentation, generative fill, the canvas workspace, and conventional editing operations. This isolated the semantic control-composition layer, but means that effects cannot be attributed to spatial persistence or segmentation alone. Future studies could compare persistent regions with conventional panels and transient selections.

Third, the current prototype provided limited visibility into why controls and initial values were proposed, and some interactions were constrained by parameter ranges, latency, and implementation limitations. Future work should investigate explanations, provenance, alternative semantic-to-parametric mappings, and mechanisms for modifying the structure of an AI proposal rather than only its values.


Finally, our study focused on interaction rather than the aesthetic quality of final images. Future work could combine interaction measures with expert evaluation or measures of alignment with users' intentions, and examine how semantic-to-parametric mappings develop through longer-term use.

Taken together, Atelier suggests that AI can support domain-specific tools by surfacing contextually relevant operations rather than replacing them. Such translations should remain provisional, revisable, and legible, giving users both the ability to manipulate AI proposals and sufficient authority over how those proposals are formed.

\section{Conclusion} 
\label{sec:conclusion} 
We introduced Atelier, a semantic interface generation approach that translates high-level image-editing intentions into context-specific, manipulable controls. Rather than producing only a finished edit, Atelier externalizes the AI's interpretation as widgets that users can inspect, refine, combine, or reject. In a within-subjects study with 12 image-editing practitioners, participants strongly preferred AI-composed controls and reported higher Exploration and Enjoyment. AI composition also increased manual editing. Participants performed substantially more interactions and parameter adjustments while contributing most edits themselves. Yet they felt significantly less in control, showing that editable AI proposals do not necessarily preserve felt control when the AI determines which actions become available and salient. Together, our findings position semantic interface generation as a mixed-initiative process in which AI helps construct the space of possible actions. Future interfaces should therefore allow users to refine not only the resulting edits, but also the AI's interpretation of what should be manipulated.

\section{AI Usage Disclosure}
The authors used generative AI tools (Claude Code, Claude Opus 4.8, and ChatGPT 5.6) to support system implementation and debugging, assist with the preparation and revision of figures and \LaTeX{} tables, and review parts of the manuscript for potential readability issues. All AI-generated outputs and suggestions were reviewed and verified by the authors.


\bibliographystyle{ACM-Reference-Format}
\bibliography{sample-base}

\appendix


\section{Implementation}

\subsection{Executing Parametric Edits With WebGL} The AI models do not directly perform ordinary adjustments such as exposure, contrast, white balance, HSL, or curves. Instead, once Claude has selected operations and proposed parameter values, these values are written into Atelier's shared editing state. The frontend then executes the corresponding operations through its WebGL rendering pipeline. The registry links each editing operation to its valid parameters, interface representation, and rendering implementation, allowing an AI-set value and a manually set value to produce the same deterministic image operation. Thus, Claude acts primarily as a semantic decision-making layer, while WebGL remains the actual engine that computes and renders the parametric image transformations.

\subsection{Runtime Stack}
The frontend is a single-page React app with no server-side rendering; it runs from one build either in the browser against a local backend or as a Vercel static bundle against a remote backend.
The backend is a FastAPI app whose Pydantic schemas are the single source of truth for the wire format and the generated TypeScript types.
\autoref{tab:app-runtime} lists both stacks with the numeric version pins from the snapshot.

\begin{table*}[htbp]
    \centering
    \caption{Frontend and backend runtime stack used to implement Atelier.}
    \Description{Frontend and backend technologies used to implement Atelier. The frontend is implemented in TypeScript and React with Vite, Zustand, and MobileSAM-based segmentation. The backend is implemented in Python with FastAPI, Uvicorn, and Pydantic, and provides access to large language models, optional server-side segmentation, and generative fill through FLUX Fill Pro.}
    \label{tab:app-runtime}

    \renewcommand{\arraystretch}{1.08}

    \begin{tabularx}{\textwidth}{
        @{}
        >{\raggedright\arraybackslash}p{0.20\textwidth}
        >{\raggedright\arraybackslash}X
        >{\raggedright\arraybackslash}X
        @{}
    }
        \toprule
        \textbf{Concern} &
        \textbf{Frontend} &
        \textbf{Backend} \\
        \midrule

        Language &
        TypeScript 5.9.3 (strict) &
        Python 3.11+ \\

        UI / web framework &
        React 19.2.0 &
        FastAPI 0.115.0 \\

        Build / ASGI server &
        Vite 7.3.1 &
        Uvicorn 0.31.0 \\

        State / validation &
        Zustand 5.0.11 &
        Pydantic 2.9.2 (S\&oT) \\

        \addlinespace[4pt]
        \multicolumn{3}{@{}l}{\textbf{AI and image services}} \\

        LLM SDK &
        --- &
        anthropic 0.39.0; claude-opus-4-7 \\

        Segmentation &
        MobileSAM ONNX via onnxruntime-web 1.26.0 (deployed) &
        opt-in server SAM 2 (\texttt{ANALYZE\_SAM=1}) \\

        Gen-fill &
        --- &
        Replicate FLUX Fill Pro over HTTPS \\

        \bottomrule
    \end{tabularx}
\end{table*}












The deployed Docker image installs the slim manifest from \texttt{pyproject.toml} and runs segmentation in the browser; the full \texttt{requirements.txt} adds \texttt{torch}, \texttt{torchvision}, and SAM~2 for the local server-side path only.

\section{Task Stimuli and Scoring}
\label{app:ms-stimuli}

\begin{figure}[htbp]
\centering
\begin{subfigure}{0.48\textwidth}
  \centering
  \includegraphics[width=\linewidth]{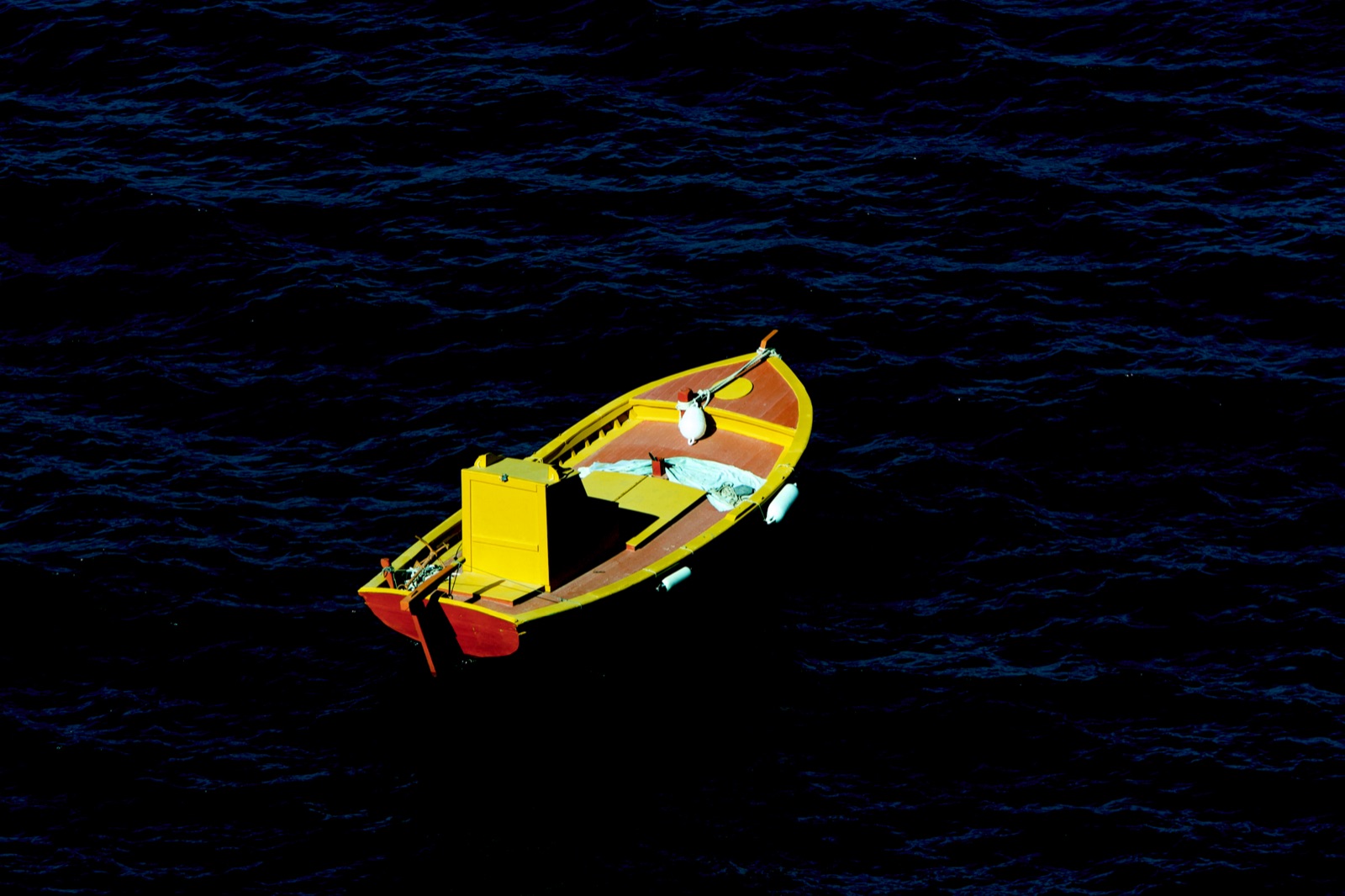}
  \caption{Boat: too dark and too cold (under-cold).}
\end{subfigure}
\hfill
\begin{subfigure}{0.48\textwidth}
  \centering
  \includegraphics[width=\linewidth]{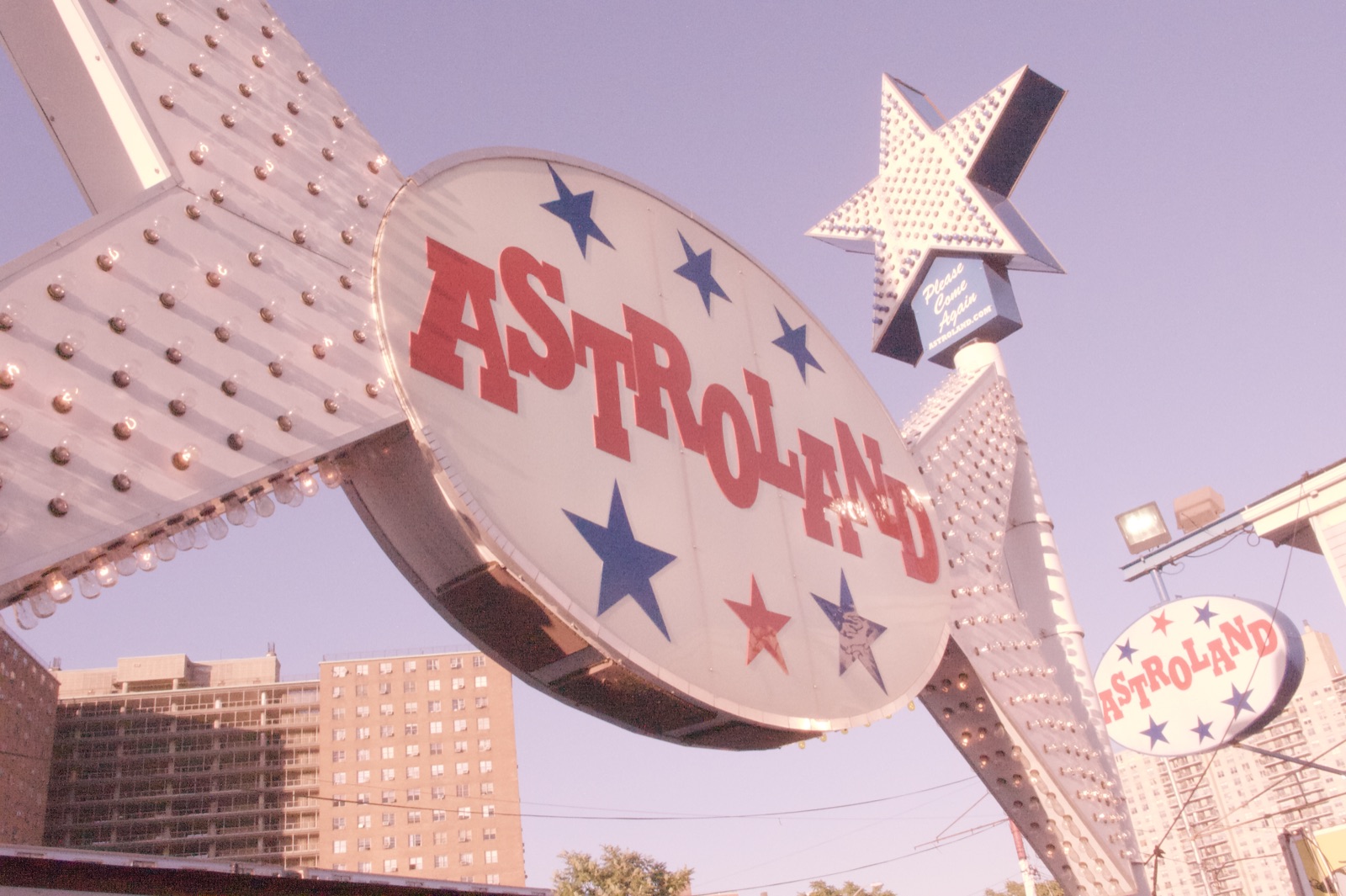}
  \caption{Sign: too bright and too warm (over-warm).}
\end{subfigure}
\caption[The two main-study stimuli.]{The two participant-facing stimuli, one edited per block: a boat degraded under-cold and a fairground sign degraded over-warm, both single saturated-color subjects from the MIT-Adobe FiveK raw dataset~\cite{bychkovsky_fivek_2011}. Each defect is baked into the DNG and recoverable from the retained original.}
\Description{Two side-by-side photos, one per stimulus. Left: a small yellow rowboat with red trim, photographed from above, floating alone on dark navy water; a white bird stands on its bow. Right: a red-and-blue ASTROLAND oval marquee sign studded with light bulbs, mounted on a star-shaped light structure, photographed from below against a blue sky, with a second smaller Astroland sign and apartment buildings visible in the background.}
\label{fig:study-stimuli}
\end{figure}

The three participant-facing briefs, verbatim (the sign versions state the mirror-image defect and its own background):
\begin{enumerate}\setlength{\itemsep}{0pt}
\item \emph{Corrective:} This photo came out of the camera too dark and too cold. Adjust it so the whole scene looks natural and true to life, the way it would have looked if you were standing there.
\item \emph{Region under protection:} The boat is still too dark. Brighten the boat so its color reads clearly, without changing the sea.
\item \emph{Open:} To finish, make the photo your own. Explore what the editor offers and take the image toward a final look you would be happy to publish. There is no right answer.
\end{enumerate}

The two photographs (Figure \ref{fig:study-stimuli}) are degraded from their retained originals by a fixed, invertible recipe baked into the DNG raw file: an exposure curve and a white-balance shift applied in opposite directions across the pair, so the boat comes out too dark and cool and the sign too bright and warm.
Working from the raw file lets the prototype's own raw-development path reproduce the defect exactly, and because the recipe is invertible the degradation is recoverable.



\section{Model Prompts}
\label{chap:appendix-prompts}

This part includes verbatim the hand-authored prompts the backend sends to the vision-language model (Anthropic's Claude) in the widget pipeline: an agent-loop \emph{system} prompt, a \emph{planner} prompt, and two \emph{resolver} prompts.
Interpolated slots are shown in braces (e.g.\ \texttt{\{targets\}}); the accompanying user message carries image context, op schemas, and starting parameters and is assembled per call.

\subsection{Agent-Loop System Prompt}
\label{app:prompt-system}

The static base, extended at runtime with extra paragraphs when the request carries pre-selected targets, pinned objects, or reference images.

\begin{lstlisting}[language={},basicstyle=\footnotesize\ttfamily,breaklines=true,breakindent=0pt]
You are an editing agent for a photo editor. The user gives an editing
request; you fulfil it by CALLING TOOLS, not by replying in prose.

To apply any tonal/colour adjustment (contrast, warmth, exposure, mood, etc.)
you MUST call propose_adjustment_widgets with target_image_node_id set to an
existing node id ({targets}) and a short intent describing the change. To put
an object on its own layer first, call copy_object_to_image_node, then
propose_adjustment_widgets on the image_node_id it returns. Do not stop until
you have called at least one tool that satisfies the request.
\end{lstlisting}

\noindent The slot \texttt{\{targets\}} is the comma-joined list of node ids, or \enquote{the active image node} when none is supplied. Optional appended blocks cover forced targets, pinned mask ids, and reference nodes the model must not edit.

\subsection{Planner Prompt}
\label{app:prompt-planner}

Fully static; user intent and image context arrive as the user message.

\begin{lstlisting}[language={},basicstyle=\footnotesize\ttfamily,breaklines=true,breakindent=0pt]
You are a photo-editing composition planner.

Given a user intent and image context, return a stack of 1-6 conceptually-grouped
photo-editing widgets. Each widget can carry 1-5 raw ops that belong together
conceptually. Each widget becomes ONE card on the user's canvas they can
independently refine.

Rules:
- A widget is ONE perceptual intention the user would name out loud
  ("lift the shadows", "warm it up"). Group ops into the same widget only
  when a user would refine them together as a single unit. The `category`
  field is a strong default: ops with the same category usually belong
  together unless you have a specific reason to split.
  BAD grouping: levels (tonal fade) + grain (texture) in one widget just
  because both feel "vintage" - a user refining grain does not expect the
  blacks to move.
- Prefer the MINIMUM set of ops that achieves the intent. Never add a
  second op that pushes the same perceptual axis as one already planned
  (e.g. light exposure AND levels brightening) - overlapping ops multiply
  and overshoot.
- Give each widget a short, descriptive `widget_name` (2-4 words) describing
  the EFFECT, not the op (e.g. "Lifted blacks", not "Levels op").
- Give each widget a `driver_label`: a 1-2 word noun naming the INTENT AXIS
  its strength slider will control ("Blackness", "Warmth", "Drama") - the
  quality the user asked for, not an op name. The frontend renders one
  slider with this label that scales the whole widget from "as shot" (0)
  to your resolved values (100).
- Prefer raw ops over presets unless the intent matches a preset closely.
- You may unfold a preset's ops as starting points and modify them.
- Order widgets by intent priority (most defining effect first).
- Emit the plan via the `emit_plan` tool.

Example for "vintage film" (abridged):
{"plan": [
  {"widget_name": "Lifted blacks", "driver_label": "Faded blacks", "ops": [...]},
  {"widget_name": "Warm fade",     "driver_label": "Warmth",       "ops": [...]},
  {"widget_name": "Film grain",    "driver_label": "Grain",        "ops": [...]}
]}
\end{lstlisting}

\subsection{Resolver Prompt (Single Operation)}
\label{app:prompt-resolver-op}

\begin{lstlisting}[language={},basicstyle=\footnotesize\ttfamily,breaklines=true,breakindent=0pt]
You are resolving numeric parameter values for a single
photo-editing operation, given the user's intent and image context.

Return strict JSON matching the op's param schema. Use the starting_params as
a strong prior if provided. Do not include markdown fences.
\end{lstlisting}

\subsection{Resolver Prompt (Whole Stack)}
\label{app:prompt-resolver-stack}

The one-call variant that resolves the entire stack together, so overlapping axes share the correction budget.

\begin{lstlisting}[language={},basicstyle=\footnotesize\ttfamily,breaklines=true,breakindent=0pt]
You are resolving numeric parameter values for an ENTIRE
stack of photo-editing operations at once, given the user's intent and image
context.

The stack's COMBINED effect must achieve the intent. Budget overlapping axes:
if two ops both raise brightness (e.g. exposure and a shadows lift), split the
correction between them - never let each op independently achieve the full
intent on its own. Prefer restrained values; the user can push further.

Use each op's starting_params as a strong prior if provided. Emit one
emit_stack_params call covering every (entry_index, op_id) in the plan.
\end{lstlisting}

\section{Questionnaires}

\subsection{Trust and Control}
\label{app:ms-trust}

Two scales on a seven-point agreement scale (Strongly disagree to Strongly agree), grounded in the F4 dimension the pre-study surfaced as decisive (\autoref{subsec:prestudy-themes}) and the trust-in-automation constructs~\cite{lee_trust_2004}. The control-and-trust scale (six items) is answered in both conditions.

\emph{Control and trust (both conditions):}
\begin{enumerate}\setlength{\itemsep}{0pt}
\item I felt in control of the editing process.
\item The final image reflected my own intentions.
\item I could predict what each control would do before I used it.
\item When I did not like a change, it was easy to undo or adjust it.
\item I always understood the current state of my edit.
\end{enumerate}

\emph{AI suggestions (AI-composed block only):}
\begin{enumerate}\setlength{\itemsep}{0pt}
\item I understood why the system suggested what it did.
\item I could see what an AI suggestion would change before I accepted it.
\item It was easy to reject or adjust the AI's suggestions.
\item I trusted the AI's suggestions.
\item Working with the AI felt like a collaboration rather than something done to me.
\item The AI's behavior was predictable.
\item The AI suggestions appeared without me asking and got in the way. \emph{(reverse)}
\end{enumerate}

\subsection{Final Comparison}
\label{app:ms-final}

The final questionnaire collects the CSI comparisons~\cite{cherry_quantifying_2014} and the direct between-condition contrast below. Versions are labeled neutrally as A (first block) and B (second block); the experimenter mapped A/B back to AI-composed/Baseline from the participant log.

\begin{itemize}\setlength{\itemsep}{0pt}
\item \textbf{FC01} Overall preference: Version A / Version B / No preference.
\item \textbf{FC02} One- or two-sentence rationale (free text).
\item \textbf{FC03} For each of six dimensions, which version was higher (A much more $\ldots$ B much more): felt more in control; trusted what the tool was doing; helped explore ideas / be creative; required less mental effort; more satisfied with the final image; felt more like own work (authorship).
\item \textbf{FC04} For a \emph{corrective} edit, would you want the AI features on? Yes / No / Depends.
\item \textbf{FC05} For a \emph{creative} edit, would you want the AI features on? Yes / No / Depends.
\item \textbf{FC06} The one thing the AI version did best (free text).
\item \textbf{FC07} The one thing that most got in the way, in either version (free text).
\end{itemize}

\section{Prompt for Generating the Main Image used for Examples}
\textit{``Photorealistic macro nature photograph, 16:9 horizontal composition. Three small pale yellow-green maple leaves resting on top of a large dark olive-green palm leaf with strong long parallel ridges running diagonally from upper left toward lower right. The maple leaves are spaced apart in a loose triangular arrangement: one leaf in the upper-left quadrant, one near the lower-center, and one in the right-center. Each maple leaf has five pointed lobes, thin visible veins, a slender stem, slightly imperfect natural edges, subtle brown tips, tiny speckles, and realistic surface texture. The leaves should be relatively small compared with the frame, leaving plenty of visible palm-leaf background around them.''}


\section{Participant Details}
\label{app:participants}

\begin{table*}[h]
    \caption{Main-study participants (n=12): self-reported demographics and experience from the pre-session questionnaire. Editing skill is the self-rating.}
    \Description{Demographics and self-reported experience of the 12 main-study participants. Participants were aged 26 to 34 and included nine men and three women, with six self-identifying as intermediate and six as professional image editors. Their editing frequency ranged from a few times per year to daily, while use of AI tools ranged from rarely to constantly. Participants represented varied creative and technical roles, including photography, graphic and user experience design, retouching, videography, creative direction, and computer science.}
    \centering
    \begin{tabular}{lllllll}
        \toprule
        P & Age & Gender & Editing experience & Editing frequency & AI frequency & Profession \\
        \midrule
        P1 & 28 & Man & Intermediate & Monthly & Sometimes & Hobbyist \\
        P2 & 31 & Man & Intermediate & A few/yr & Constantly & Hobbyist \\
        P3 & 26 & Woman & Intermediate & Monthly & Sometimes & UX Designer \\
        P4 & 27 & Man & Intermediate & A few/yr & Rarely & Photographer \\
        P5 & 34 & Woman & Intermediate & Monthly & Often & Computer Scientist \\
        P6 & 30 & Woman & Intermediate & Weekly & Often & Freelance Creative Director \\
        P7 & 27 & Man & Professional & Monthly & Often & Photographer \\
        P8 & 26 & Man & Professional & Daily & Rarely & Videographer \\
        P9 & 28 & Man & Professional & Monthly & Constantly & Retoucher \\
        P10 & 28 & Man & Professional & Daily & Rarely & Graphic designer \\
        P11 & 29 & Man & Professional & Weekly & Sometimes & Art director / Freelancing Artist \\
        P12 & 27 & Man & Professional & Weekly & Often & Graphic designer \\
        \bottomrule
    \end{tabular}
    \label{table:participants}
\end{table*}


\section{Results}
\label{app:results}
\begin{table*}[htbp]
\centering
\caption{NASA-TLX: AI-composed vs.\ Baseline ($N=12$).}
\Description{NASA Raw Task Load Index scores comparing AI-composed and Baseline conditions for 12 participants. The AI-composed condition shows lower mean scores than the Baseline for mental demand, physical demand, temporal demand, effort, frustration, and overall workload, while perceived performance is identical across conditions. None of the individual workload dimensions or the overall score differ significantly between conditions.}
\label{tab:tlx}
\resizebox{\textwidth}{!}{%
\begin{tabular}{lccccccccccc}
\toprule
& \multicolumn{2}{c}{AI-composed} & \multicolumn{2}{c}{Baseline} & \multicolumn{1}{c}{AI-composed} & \multicolumn{1}{c}{Baseline} & & & \multicolumn{3}{c}{Statistics} \\
\cmidrule(lr){2-3} \cmidrule(lr){4-5} \cmidrule(lr){6-6} \cmidrule(lr){7-7} \cmidrule(lr){10-12}
Measure & $M$ & $SD$ & $M$ & $SD$ & $Mdn$ $(IQR)$ & $Mdn$ $(IQR)$ & Shapiro $p$ & Test & $p$ & Effect size & 95\% CI \\
\midrule
Mental demand & 2.67 & 1.37 & 3.42 & 1.51 & 2.50 (1.75--4.00) & 3.00 (3.00--4.25) & 0.377 & Paired $t$ & 0.191 & $d_z = -0.40$ & $[-1.93, 0.43]$ \\
Physical demand & 1.42 & 0.79 & 2.00 & 1.28 & 1.00 (1.00--1.25) & 1.50 (1.00--3.00) & 0.001 & Wilcoxon & 0.141 & $r = -0.80$ & $[-1.50, 0.00]$ \\
Temporal demand & 2.33 & 1.78 & 2.50 & 1.45 & 2.00 (1.00--2.00) & 2.00 (1.75--3.25) & 0.020 & Wilcoxon & 0.516 & $r = -0.29$ & $[-1.00, 0.50]$ \\
Performance & 5.42 & 1.00 & 5.42 & 0.79 & 5.00 (5.00--6.00) & 5.00 (5.00--6.00) & 0.011 & Wilcoxon & 1.000 & $r = 0.00$ & $[-0.50, 0.50]$ \\
Effort & 2.67 & 1.15 & 3.50 & 1.83 & 2.00 (2.00--3.25) & 3.50 (2.00--5.00) & 0.150 & Paired $t$ & 0.166 & $d_z = -0.43$ & $[-2.07, 0.40]$ \\
Frustration & 2.75 & 1.22 & 2.92 & 1.62 & 2.00 (2.00--3.25) & 2.50 (1.75--4.25) & 0.448 & Paired $t$ & 0.723 & $d_z = -0.11$ & $[-1.17, 0.84]$ \\
\midrule
Overall & 2.40 & 0.82 & 2.82 & 0.94 & 2.50 (1.79--2.87) & 2.92 (2.33--3.67) & 0.196 & Paired $t$ & 0.130 & $d_z = -0.47$ & $[-0.98, 0.14]$ \\
\bottomrule
\end{tabular}%
}
\begin{minipage}{\textwidth}
\vspace{2pt}
\footnotesize \textit{Note.} $N=12$. Test per row selected via Shapiro--Wilk on the paired differences
(AI-composed $-$ Baseline) at $\alpha=.05$: paired $t$-test (Cohen's $d_z$) where normal, Wilcoxon signed-rank
test (matched-pairs rank-biserial $r$) otherwise. CIs are on the mean difference for paired-$t$ rows and
the Hodges--Lehmann estimate of the median difference (bootstrap, 10{,}000 resamples) for Wilcoxon rows.
\end{minipage}
\end{table*}


\begin{table*}[htbp]
\centering
\caption{Control / Trust: AI-composed vs.\ Baseline ($N=12$).}
\Description{Control and trust ratings comparing AI-composed and Baseline conditions for 12 participants. Participants reported significantly lower feelings of control in the AI-composed condition than in the Baseline. Ratings for predictability, ease of undoing or adjusting changes, and understanding the current state were also lower for AI-composed, but these differences were not statistically significant. Ratings of how well the result reflected participants’ intentions were nearly identical across conditions.}
\label{tab:trust}
\resizebox{\textwidth}{!}{%
\begin{tabular}{lccccccccccc}
\toprule
& \multicolumn{2}{c}{AI-composed} & \multicolumn{2}{c}{Baseline} & \multicolumn{1}{c}{AI-composed} & \multicolumn{1}{c}{Baseline} & & & \multicolumn{3}{c}{Statistics} \\
\cmidrule(lr){2-3} \cmidrule(lr){4-5} \cmidrule(lr){6-6} \cmidrule(lr){7-7} \cmidrule(lr){10-12}
Measure & $M$ & $SD$ & $M$ & $SD$ & $Mdn$ $(IQR)$ & $Mdn$ $(IQR)$ & Shapiro $p$ & Test & $p$ & Effect size & 95\% CI \\
\midrule
Felt in control & 5.25 & 1.22 & 5.83 & 1.03 & 5.50 (5.00--6.00) & 6.00 (5.75--6.25) & 0.133 & Paired $t$ & \textbf{0.046} & $d_z = -0.65$ & $[-1.16, -0.01]$ \\
Result reflected intentions & 5.83 & 1.27 & 5.83 & 0.58 & 6.00 (5.75--7.00) & 6.00 (5.75--6.00) & 0.011 & Wilcoxon & 0.942 & $r = -0.03$ & $[-0.50, 0.50]$ \\
Could predict each control & 4.42 & 1.44 & 4.75 & 1.06 & 4.50 (4.00--5.00) & 5.00 (4.00--5.25) & 0.160 & Paired $t$ & 0.220 & $d_z = -0.38$ & $[-0.90, 0.23]$ \\
Easy to undo/adjust & 5.17 & 1.64 & 6.00 & 1.13 & 5.00 (3.75--7.00) & 6.00 (5.75--7.00) & 0.023 & Wilcoxon & 0.054 & $r = -0.79$ & $[-1.50, 0.00]$ \\
Understood current state & 5.08 & 1.31 & 5.42 & 1.38 & 5.00 (4.75--6.00) & 5.50 (5.00--6.25) & 0.035 & Wilcoxon & 0.279 & $r = -0.48$ & $[-1.00, 0.00]$ \\
\bottomrule
\end{tabular}%
}
\begin{minipage}{\textwidth}
\vspace{2pt}
\footnotesize \textit{Note.} $N=12$. Test per row selected via Shapiro--Wilk on the paired differences
(AI-composed $-$ Baseline) at $\alpha=.05$: paired $t$-test (Cohen's $d_z$) where normal, Wilcoxon signed-rank
test (matched-pairs rank-biserial $r$) otherwise. CIs are on the mean difference for paired-$t$ rows and
the Hodges--Lehmann estimate of the median difference (bootstrap, 10{,}000 resamples) for Wilcoxon rows.
\end{minipage}
\end{table*}


\begin{figure*}[htbp]
\centering
\definecolor{seriesblue}{HTML}{2A78D6}
\definecolor{seriesorange}{HTML}{EB6834}
\begin{tikzpicture}
\begin{axis}[
    width=12cm, height=4.2cm,
    xmin=0, xmax=100,
    ymin=0.4, ymax=2.6,
    xtick={0,20,...,100},
    ytick={1,2},
    yticklabels={AI-composed, Baseline},
    xlabel={SUS Score},
    xmajorgrids=true,
    grid style={gray!25},
    axis lines*=left,
    axis line style={gray!60},
    tick align=outside,
    tick style={gray!60},
    y axis line style={draw=none},
    ytick style={draw=none},
    boxplot/draw direction=x,
    boxplot/box extend=0.35,
    boxplot/whisker range=1.5,
    boxplot/every box/.style={draw=black, semithick},
    boxplot/every whisker/.style={black, densely dashed},
    boxplot/every median/.style={black, thick},
    boxplot/every outlier/.style={
        mark=o, mark size=2pt, draw=black, fill=none, only marks,
    },
]

\addplot+[boxplot, boxplot/draw position=2, fill=seriesorange!30, draw=seriesorange!90!black]
    table[row sep=\\, y index=0] {
    data \\
    72.5 \\
    72.5 \\
    90.0 \\
    60.0 \\
    80.0 \\
    42.5 \\
    57.5 \\
    82.5 \\
    75.0 \\
    67.5 \\
    85.0 \\
    75.0 \\
    };

\addplot+[boxplot, boxplot/draw position=1, fill=seriesblue!25, draw=seriesblue!90!black]
    table[row sep=\\, y index=0] {
    data \\
    82.5 \\
    55.0 \\
    82.5 \\
    67.5 \\
    77.5 \\
    70.0 \\
    75.0 \\
    82.5 \\
    70.0 \\
    70.0 \\
    82.5 \\
    77.5 \\
    };

\end{axis}
\end{tikzpicture}
\caption{System Usability Scale (SUS) results comparing Baseline and AI-composed ($N=12$).}
\Description{Horizontal boxplot comparing System Usability Scale scores
between two conditions, N=12, on an axis from 0 to 100 labeled SUS Score. Baseline has a median around 74, an interquartile range from about 64 to 81, and whiskers extending from 42.5 to 90, with no outliers. AI-composed has a median around 76, a narrower interquartile range from 70 to 82.5, and a whisker extending down to 55; it has no upper whisker, since its highest score, 82.5, sets the box's own upper edge. Overall, AI-composed scores are similar to or slightly higher than Baseline, and more tightly clustered.}
\label{fig:sus-boxplot}
\end{figure*}
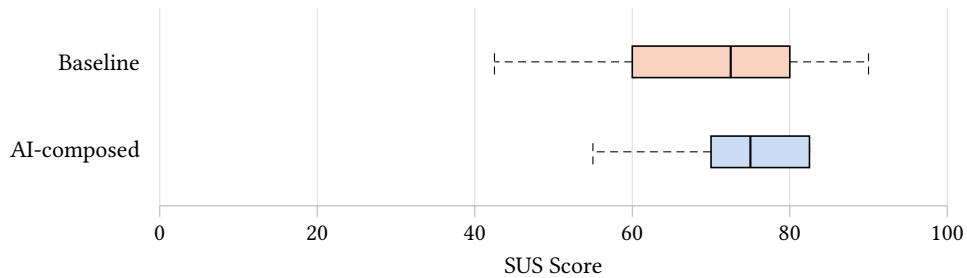


\begin{table*}[t]
\centering
\caption{Exploratory comparison of CSI outcomes by expertise
($n=6$ intermediate, $n=6$ professional). Condition scores are reported
as mean (SD), and within-participant changes
($\Delta=\text{AI-composed}-\text{Baseline}$) as median [IQR].
Positive $\Delta$ values indicate greater creativity support with AI.
Between-tier differences in change scores were tested using two-sided
Mann--Whitney $U$ tests. Factor-level $p$-values are unadjusted and
exploratory.}
\Description{Exploratory comparison of Creativity Support Index outcomes for intermediate and professional participants. Both expertise groups show higher overall Creativity Support Index scores with AI-composed than with Baseline, with a larger median increase among professionals. Professionals also show larger positive changes in expressiveness and exploration, while intermediates show larger changes in immersion and results worth effort. None of the differences in change scores between expertise groups are statistically significant.}
\label{tab:csi-expertise}
\footnotesize
\setlength{\tabcolsep}{4pt}

\resizebox{\textwidth}{!}{%
\begin{tabular}{@{}lccccccc@{}}
\toprule
&
\multicolumn{2}{c}{\textbf{Intermediate}}
&
\multicolumn{2}{c}{\textbf{Professional}}
&
\multicolumn{3}{c}{\textbf{Expertise difference}} \\
\cmidrule(lr){2-3}
\cmidrule(lr){4-5}
\cmidrule(l){6-8}

\textbf{Measure}
& \shortstack{\textbf{Baseline/ AI-composed}\\\textbf{$M$ $(SD)$}}
& \shortstack{\textbf{$\Delta$}\\\textbf{$Mdn$ $[IQR]$}}
& \shortstack{\textbf{Baseline / AI-composed}\\\textbf{$M$ $(SD)$}}
& \shortstack{\textbf{$\Delta$}\\\textbf{$Mdn$ $[IQR]$}}
& $\boldsymbol{U}$
& $\boldsymbol{p}$
& $\boldsymbol{\delta}$ \\
\midrule

Overall CSI
& 74.17 (18.33) / 77.83 (13.13)
& 1.75 [$-7.75$, 7.13]
& 73.30 (16.24) / 82.83 (8.35)
& 12.35 [1.93, 15.13]
& 11.5 & .336 & .36 \\

Expressiveness
& 14.33 (6.22) / 15.00 (4.20)
& $-0.50$ [$-1.00$, 0.75]
& 13.50 (3.62) / 17.17 (1.47)
& 4.00 [0.25, 6.25]
& 10.5 & .252 & .42 \\

Exploration
& 14.50 (5.09) / 16.00 (3.29)
& 0.50 [$-1.75$, 2.00]
& 13.00 (4.69) / 17.17 (3.06)
& 4.50 [2.25, 6.00]
& 9.0 & .169 & .50 \\

Immersion
& 10.67 (4.84) / 12.83 (4.79)
& 3.00 [1.25, 5.50]
& 14.33 (3.61) / 13.67 (3.50)
& 0.50 [$-2.50$, 2.75]
& 24.5 & .334 & $-.36$ \\

Enjoyment
& 15.17 (3.43) / 16.33 (3.14)
& 0.50 [$-0.75$, 2.50]
& 16.00 (3.52) / 18.17 (1.33)
& 1.00 [0.00, 2.75]
& 14.0 & .567 & .22 \\

Results worth effort
& 15.33 (3.08) / 16.17 (3.31)
& 2.00 [$-1.75$, 3.50]
& 16.00 (2.10) / 16.17 (2.32)
& 0.50 [0.00, 1.75]
& 22.5 & .517 & $-.25$ \\

\bottomrule
\end{tabular}%
}
\end{table*}

\begin{table*}[t]
\centering
\caption{Exploratory comparison of NASA-TLX outcomes by expertise
($n=6$ intermediate, $n=6$ professional). Condition scores are reported
as mean (SD), and within-participant changes
($\Delta=\text{AI-composed}-\text{Baseline}$) as median [IQR].
Negative $\Delta$ values indicate reduced workload with AI assistance.
Performance was reverse-scored so that higher values indicate greater
workload. Factor-level $p$-values are unadjusted and exploratory.}
\Description{Exploratory comparison of NASA Raw Task Load Index outcomes for intermediate and professional participants. Intermediate participants generally show larger reductions in workload with AI-composed than professionals, particularly for mental demand and effort. The expertise groups differ significantly only in the change in mental demand, with intermediates showing a larger reduction. Differences between expertise groups for overall workload and the remaining workload dimensions are not statistically significant.}
\label{tab:rtlx-expertise}
\footnotesize
\setlength{\tabcolsep}{4pt}

\resizebox{\textwidth}{!}{%
\begin{tabular}{@{}lccccccc@{}}
\toprule
&
\multicolumn{2}{c}{\textbf{Intermediate}}
&
\multicolumn{2}{c}{\textbf{Professional}}
&
\multicolumn{3}{c}{\textbf{Expertise difference}} \\
\cmidrule(lr){2-3}
\cmidrule(lr){4-5}
\cmidrule(l){6-8}

\textbf{Measure}
& \shortstack{\textbf{Baseline/ AI-composed}\\\textbf{$M$ $(SD)$}}
& \shortstack{\textbf{$\Delta$}\\\textbf{$Mdn$ $[IQR]$}}
& \shortstack{\textbf{Baseline / AI-composed}\\\textbf{$M$ $(SD)$}}
& \shortstack{\textbf{$\Delta$}\\\textbf{$Mdn$ $[IQR]$}}
& $\boldsymbol{U}$
& $\boldsymbol{p}$
& $\boldsymbol{\delta}$ \\
\midrule

Overall TLX
& 3.17 (0.84) / 2.36 (0.93)
& $-0.59$ [$-1.34$, $-0.21$]
& 2.47 (0.97) / 2.44 (0.79)
& 0.42 [$-0.55$, 0.50]
& 7.5 & .106 & .58 \\

Mental demand
& 4.50 (1.05) / 2.50 (1.38)
& $-2.50$ [$-3.00$, $-0.50$]
& 2.33 (1.03) / 2.83 (1.47)
& 0.50 [0.00, 1.00]
& 4.0 & .027 & .78 \\

Physical demand
& 2.33 (1.51) / 1.50 (0.84)
& 0.00 [$-1.50$, 0.00]
& 1.67 (1.03) / 1.33 (0.82)
& 0.00 [0.00, 0.00]
& 17.0 & .924 & .06 \\

Temporal demand
& 2.33 (1.51) / 2.17 (1.94)
& 0.00 [$-0.75$, 0.75]
& 2.67 (1.51) / 2.50 (1.76)
& 0.00 [0.00, 0.00]
& 18.5 & 1.000 & $-.03$ \\

Performance demand
& 2.83 (0.41) / 3.00 (1.10)
& 0.00 [0.00, 0.75]
& 2.33 (1.03) / 2.17 (0.75)
& 0.00 [0.00, 0.00]
& 21.0 & .652 & $-.17$ \\

Effort
& 3.67 (1.75) / 2.33 (1.03)
& $-1.00$ [$-2.50$, $-0.25$]
& 3.33 (2.07) / 3.00 (1.26)
& 0.00 [0.00, 0.75]
& 9.0 & .157 & .50 \\

Frustration
& 3.33 (1.63) / 2.67 (1.21)
& $-0.50$ [$-1.75$, 0.00]
& 2.50 (1.64) / 2.83 (1.33)
& 0.50 [0.00, 1.00]
& 11.0 & .287 & .39 \\

\bottomrule
\end{tabular}%
}
\end{table*}


\begin{table*}[htbp]
\centering
\caption{Supplemental data for \autoref{fig:results-fc}: response counts per
category for the final comparison ($N=12$). Categories run from strongly
favors Baseline to strongly favors AI-composed.}
\Description{Distribution of participants’ final comparison responses between AI-composed and Baseline. Eleven of 12 participants preferred the AI-composed version, while one reported no preference. Responses also tend to favor AI-composed for exploring ideas and creativity, reducing mental effort, and satisfaction with the resulting image. Perceived control and trust are more mixed, while sense of authorship predominantly favors the Baseline or is rated neutral.}
\label{tab:results-fc-data}
\small
\begin{tabular}{lccccc}
\toprule
Measure & Strongly Baseline & Baseline & Neutral & AI-composed & Strongly AI-composed \\
\midrule
Preferred version          & 0 & 0 & 1 & 11 & 0 \\
Felt in control            & 2 & 4 & 3 & 1  & 2 \\
Trusted the tool           & 1 & 3 & 5 & 1  & 2 \\
Explore ideas / creativity & 0 & 0 & 4 & 6  & 2 \\
Less mental effort         & 0 & 3 & 3 & 2  & 4 \\
Satisfied with image       & 1 & 1 & 4 & 4  & 2 \\
Sense of authorship        & 2 & 2 & 7 & 1  & 0 \\
\bottomrule
\end{tabular}
\end{table*}




\end{document}